\documentclass[aps,titlepage,floatfix,longbibliography]{revtex4-2}

\usepackage{xr}
\usepackage{hyperref}
\usepackage{amsmath}
\usepackage{physics}
\usepackage[pdftex]{graphicx}
\usepackage{xcolor}
\usepackage[papersize={8.5in,11in}]{geometry}

\begin{document}


\title{Collective gene dynamics leave signatures of decision landscapes in cell fate coordinates}

\author{Maria Yampolskaya}
\email{mariay@bu.edu}
\affiliation{Department of Physics, Boston University, Boston, MA 02215, USA}
\author{Laertis Ikonomou}
\affiliation{Department of Oral Biology, University at Buffalo School of Dental Medicine, Buffalo, NY 14215, USA }
\affiliation{Cell, Gene and Tissue Engineering Center, University at Buffalo, Buffalo, NY 14260, USA}
\affiliation{Chemical and Biological Engineering Department, University at Buffalo, Buffalo, NY 14260, USA}
\author{Pankaj Mehta}
\email{pankajm@bu.edu}
\affiliation{Department of Physics, Boston University, Boston, MA 02215, USA}
\affiliation{Center for Regenerative Medicine of Boston University and Boston Medical Center, Boston, MA 02215, USA}
\affiliation{Faculty of Computing and Data Science, Boston University, Boston, MA 02215, USA}
\affiliation{Biological Design Center, Boston University, Boston, MA 02215, USA}

\begin{abstract}
Multicellular organisms contain a wide variety of highly specialized cell types. The consistency and robustness of developmental trajectories suggest that complex gene regulatory networks effectively act as low-dimensional cell fate landscapes. Prior work inspired by dynamical systems theory argues that cell fate transitions fall into universal decision-making classes, but the theory connecting these geometric landscapes to high-dimensional gene expression space is still in its infancy. Here, we introduce a phenomenological model that identifies experimental signatures of decision-making classes in single-cell RNA-sequencing time-series data. The model combines low-dimensional gradient-like dynamics with high-dimensional Hopfield networks to capture the interplay between cell fate, gene expression, and signaling. We apply the framework to experimental mouse data on maturing lung alveolar cells and lineage-traced hematopoietic differentiation and show that the measured cell fate dynamics are consistent with developmental landscapes containing intermediate progenitors and saddle points. We further show that the framework can be used to understand spatial patterning and cell fate organization, focusing on Notch signaling in lung airways. Together, these results provide evidence that collective transcriptomic dynamics carry signatures of landscape features associated with universal decision-making classes.
\end{abstract}
\maketitle
\maketitle

\section{Introduction}
Many animals have complex organs with specialized cells that coordinate to maintain homeostasis. Cells differentiate into these cell types over the course of embryonic development. In each cell, thousands of genes work together and receive information from chemical signals and mechanical cues to determine the cell's lineage and function \cite{wolpert2015principles}. Although gene regulation is complex and high-dimensional, cells follow remarkably consistent developmental programs \cite{debat2001mapping,wagner2013robustness}. The ways cells transition between types is robust: even when perturbed, they tend towards these development trajectories. Differentiated cells are stable in their cell fate unless exposed to particular signals. In injury and in experimental protocols specially designed to induce transdifferentiation, even mature adult cells can change their type \cite{takahashi2006induction,takahashi2016decade,srivastava2016vivo,shivdasani2021tissue}. An open problem in cell biology is understanding how cells retain and transition between cell fates in development, injury, and directed differentiation. 

The experimental approach to understanding cell fate transitions often involves probing a cell's gene expression profile \cite{trapnell2015defining}. Gene expression is a major factor in determining which proteins are created, so observing gene expression gives information about a cell's function and therefore its type. Single-cell RNA-sequencing (scRNA-seq) is a widely-used technique for measuring gene expression profiles by capturing RNA from individual cells \cite{kolodziejczyk2015technology,svensson2018exponential}. There is an abundance of scRNA-seq data describing cell type, often in the form of scRNA-seq atlases that span entire organisms \cite{jones376tabula, han2018mapping, qiu2024single}. Despite the quantity of data, the theory of cell fate transitions is still in its infancy \cite{lahnemann2020eleven}. Additionally, although definitions of cell fate, identity, and type can vary, for the purposes of this paper we use them interchangeably to mean the functional and morphological cellular phenotypes that arise in typical development.

One of the earliest theoretical efforts to understand these transitions is the Waddington landscape, a metaphor for conceptualizing differentiation \cite{waddington2014strategy}. Because cell types are robust and discrete, Waddington proposed that cell fates act like attracting basins in a landscape. The developing cell is represented by a ball rolling down valleys to end up in one such basin. This metaphor suggests that the complex process of differentiation is effectively a low-dimensional landscape. The formalization of the Waddington landscape has been a long-standing problem, often approached from the perspective of dynamical systems \cite{huang2012molecular,thom2018structural,wang2011quantifying,ferrell2012bistability,moris2016transition,saez2022statistically,wang2022perspectives}. In the language of dynamical systems, the Waddington landscape says that cell types act like attractor states: stable steady states which draw in nearby points. 

Building on this work, \citet{rand2021geometry} argue that we can categorize cell fate decisions and their landscapes according to general qualitative classes defined by bifurcations. These classes are characterized by the numbers of attractors, saddle points (which, in three dimensions, can be thought of as hills between attractor basins) and the paths connecting these objects. These classes are generic: any dynamical system with an associated landscape and attractor states can be described at least locally by these decision-making classes. Identifying classes of transitions is powerful because it opens the door to finding common and universal features across transitions, a deeper understanding of how cell fates relate to one another, and the ability to predict transitions based on general principles.

This framework has been successfully applied to find the landscapes for a variety of cell fate decisions, from early mesodermal specification in mice to vulval differentiation in \textit{C. elegans}.   \cite{raju2023geometrical,rand2021geometry,camacho2021quantifying,saez2022statistically}. Additionally, \citet{cislo2025reconstructing}
 and \citet{mochulska2025generative} provide robust algorithms for inferring landscapes from flow cytometry, scRNA-seq, and cell type proportion data under varying signaling conditions. These works provide evidence that the construction of landscapes lends insight into the topology (which transitions are possible) and topography (for example, relative placement of features and attractor basin size) underlying specific systems. 

While these works have established the validity of the landscape approach, mapping between landscapes and dynamics remains an open problem. It is known that landscapes, or potential functions, do not fully specify the resulting dynamics \cite{saez2022dynamical,mochulska2025generative}. The aforementioned papers posit different ways of translating experimental data into low-dimensional dynamics, and then interpreting those dynamics in terms of a landscape. For the first step of extracting cell identity from flow cytometry or scRNA-seq data, they take one of two approaches: defining discrete cell states and observing proportions of cell types over time, or identifying a narrow number of marker genes and defining a manifold on these genes. In the case of discrete variables, valuable information about mixed states or multilineage expression is lost. In the second approach, identifying a manifold on a small number of lineage-specific marker genes, or applying dimensional reduction algorithms on these genes, results in cell state variables that are highly dependent on the system.

While previous approaches have been highly effective for constructing comprehensive landscapes for specific systems, it is difficult to scale them to study the large amount of data in scRNA-seq atlases, which regularly feature hundreds of cell types. With the aim of finding universal properties of cell fate trajectories across contexts, we introduce a mapping from high-dimensional gene expression to landscape-based cell fate dynamics. Our approach is both systematic (easily applicable to any system with the use of annotated cell type information, which is widely available) and uses continuous variables, revealing mixed states that would not be detected by discrete definitions of cell type. The core of this approach lies in defining concrete cell fate coordinates (first introduced in \citet{yampolskaya2023sctop}) and then defining dynamics on these coordinates based on different landscapes. Regardless of the specific context, we predict that the cell fate dynamics on these coordinates will be reflective of the topography and topology of the underlying landscape. The main strength of this approach is its generality: different landscapes can be directly interpreted from time series plots of trajectories, and trajectories of any cell fate can be directly compared. Additionally, the phenomena of multilineage expression -- where progenitor types express a combination of genes related to their eventual fates \cite{grishechkin2025mathematical,hu1997multilineage,olsson2016single} -- can be visualized as intermediate cell states having significant values in more than one of these cell type coordinates.

Inspired by the coarse-grained variables and phenomenological description of attractor networks, our model combines the pattern retrieval properties of modern Hopfield models with decision-making classes. Hopfield models are high-dimensional networks with stored patterns that act as attractor states \cite{hopfield1982neural,krotov2016dense,yampolskaya2025hopfield}. A growing body of work has developed the use of Hopfield networks as models of differentiation, with cell fates acting as the attractor states. \cite{lang2014epigenetic,fard2016not,pusuluri2017cellular,guo2017hopland,ikonomou2020vivo,yampolskaya2023sctop,smart2023emergent,karin2024enhancernet,boukacem2024waddington,grishechkin2025mathematical}. Our model relies extensively on the use of Hopfield-inspired order parameters, which provide consistent, interpretable cell fate coordinates for analyzing bulk and single-cell RNA-seq \cite{lang2014epigenetic,pusuluri2017cellular,ikonomou2020vivo,yampolskaya2023sctop,souza2026parameter}. Additionally, we introduce an extension of Hopfield networks that involves modular insertion of landscapes corresponding to different decision-making classes. Hopfield models typically have static attractors whereas actual cell fate transitions involve signals which destabilize attractors, causing bifurcations that can be described with landscapes and decision-making classes. 

To illustrate the interpretative power of our model, we predict representative trajectories corresponding to three classes of decision-making, identify qualitative features unique to each class, and posit the presence of these class features in existing time series data. In the differentiation of mouse lung epithelial cells, we find trajectories suggesting a landscape containing an intermediate progenitor. In an analysis of mouse hematopoietic lineage tracing data, we see trajectories containing features of decision classes with different configurations of saddle points between attractors. Although further work is necessary, our preliminary studies are promising indicators that our model can bridge between scRNA-seq time series data   and decision-making classes, and in doing so has the potential to universal properties of cell fate transitions.

\section{Results}

\subsection{A model of differentiation unifying the spaces of gene expression, cell fate, and signaling}


\begin{figure}[p]
  \centering
  \includegraphics[width=0.8\textwidth]{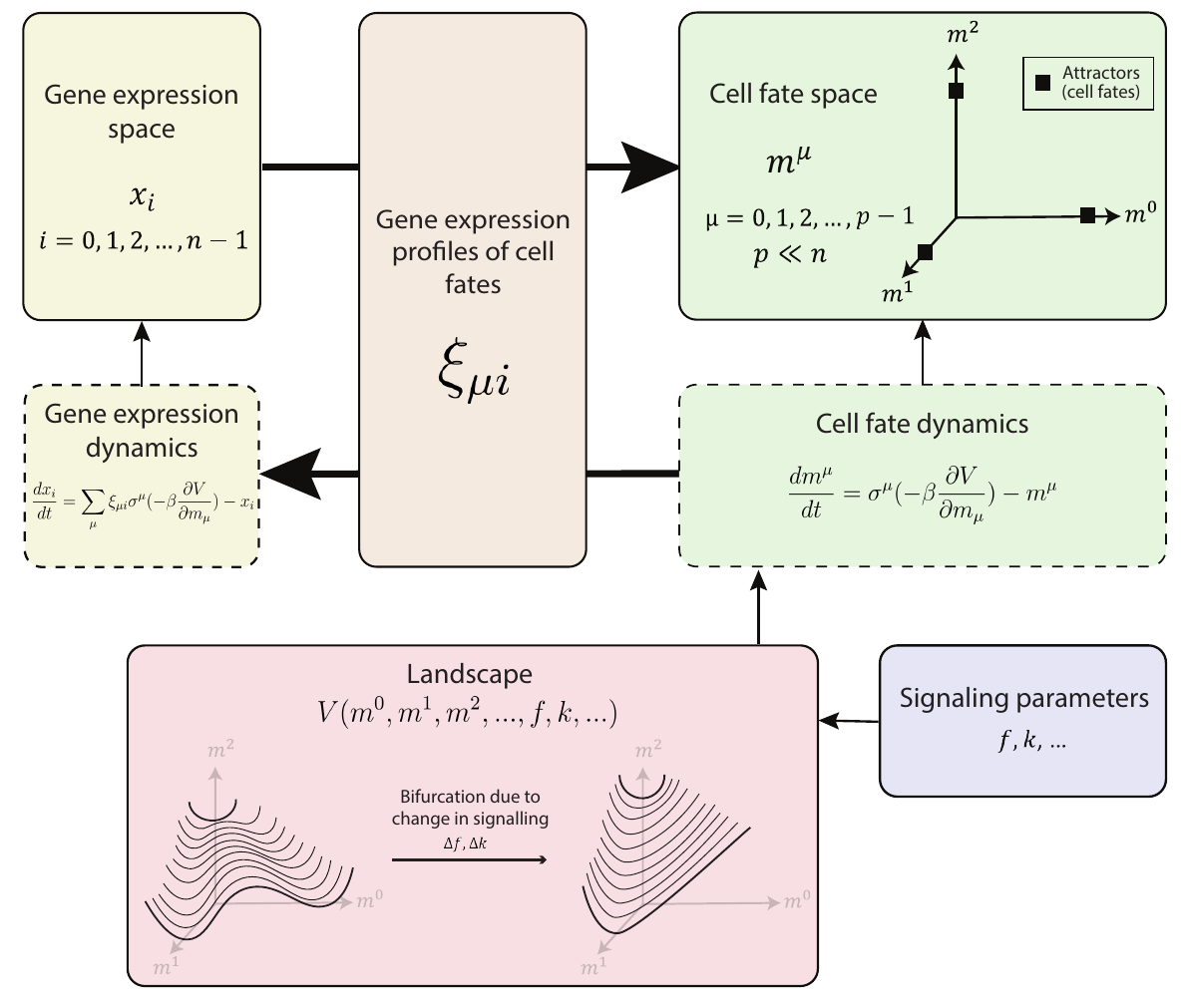}
  \caption{{\bf Overview of high-dimensional gene expression space and corresponding low-dimensional landscapes.} Each cell's state is described by gene expression $x_i$. Hopfield-inspired order parameters $m^{\mu}$ measure alignment with cell fate $\mu$, defined by reference profiles $\xi_{\mu i}$, and span a $p$-dimensional cell fate space in which fate $\mu$ sits at $m^{\mu}=1$, $m_{\nu\neq\mu}=0$. The dynamics of $x_i$ and $m^{\mu}$ exhibit Hopfield-like pattern retrieval, with cell fates as attractors, while traversing a landscape $V$ set by signaling parameters $f, k, \ldots$; changes in these parameters represent changes in signals received by the cell and can induce bifurcations.}
  \label{cartoon}
\end{figure}

In our model, cell fate is defined as a transcriptome-wide cell state that is stable under certain environmental and signaling conditions. In other words, given the requisite environmental conditions, a cell fate is an attractor in gene expression space: a stable fixed point that attracts nearby cells \cite{strogatz2001nonlinear}. Even transient attractors, like progenitor cell types in development that eventually lead to mature types, can be captured by this model given that they are temporarily stable under some specific signaling conditions. While other landscape-based models of cell fate involve abstract measures of cell fate, or system-specific quantities, we define concrete cell fate coordinates applicable to any cell fate transition, whether in development, injury, or directed differentiation. By defining a landscape on these coordinates where attractors are controlled by signaling, our model combines the three most important spaces involved: cell fate, gene expression, and signaling. Figure \ref{cartoon} summarizes the interactions between these key components. 

Our mathematical construction is based on a generalization of the Modern Hopfield network, a dynamical system for associative memory \cite{hopfield1982neural,krotov2016dense,hopfield1984neurons}. Hopfield networks have proven to be insightful models of emergent funcion across biological contexts, from memory retrieval to molecular assembly \cite{yampolskaya2025hopfield}. This paper builds on previous work connecting neural network models of associative memory with Waddington's landscape \cite{fard2016not,guo2017hopland,grishechkin2025mathematical,boukacem2024waddington}. Hopfield networks are an appealing candidate for studying cell fate landscapes because in both systems, the dynamics are defined in terms of an interacting network with prescribed attractor states. In the Hopfield model, the nodes of the network correspond to the firing rate of neurons and the attractors to stored memories. In the context of development, the nodes of our network correspond to the expression of different genes and the attractors to cell fates. We chose modern instead of classic Hopfield models because they lend themselves more naturally to continuous states and phenomenological descriptions. For example, unlike the classic model, there is no explicit interaction matrix between the nodes, and interactions of all orders are involved \cite{krotov2020large}.

Our model builds upon Modern Hopfield networks by implementing two important changes that allow us to model cell fate transitions: (1) the use of generalized order parameters and (2) the inclusion of signal-dependent forces that cause transitions between attractors \cite{yampolskaya2023controlling}. With these changes, the attractors (or memories) in our network change stability in response to external cues via signal-induced bifurcations. In the main text, we focus on giving an overview of our model. A detailed discussion of our construction can be found in the Methods and Supplemental Information, including a discussion of the relationship between this model and pure gradient systems.

In our model, the state of a cell is specified by its gene expression profile. At any time $t$, the state of the system is described by a $N$-dimensional real-valued vector $x_i(t)$ that encodes the expression of gene $i$ at time $t$.  We focus on the differentiation dynamics of a system with $p$ possible cell fates ($\mu=0,\ldots,p-1$). These cell fates are defined by a list of $p$ vectors in gene expression space, $\{\xi_{\mu i} \}$, which specify the expression level of gene $i$ in cell fate $\mu$. The $\{\xi_{\mu i} \}$ are an input to our model and can be calculated from single-cell atlases \cite{yampolskaya2023sctop,jones376tabula,han2018mapping,qiu2024single}. 

We would like a way to represent the current state of the system not only in gene expression space ($x_i(t)$), but also in the space of possible cell fates \cite{yampolskaya2023sctop}. Given a set of cell fates $\{\xi_{\mu i} \}$, we can define two different quantities that map the gene expression state $x_i(t)$ to a $p$-dimensional vector in cell fate space. The first is through defining the overlap, which is equivalent to magnetization in spin systems:
\begin{align}
m_{\mu}(t)= \frac{1}{N} \sum_j \xi_{\mu j} x_j(t).
\label{Eq:defmag}
\end{align}
As can be seen from the expression above,  the $m_{\mu}(t)$ are simply the dot products of the current gene expression state with the expression profiles of each of the $p$ cell fates we wish to model. While informative, one major drawback of $m_\mu$ is that if two cell fates are highly correlated (as is often the case for closely related lineages) the corresponding magnetizations will also be similar.

For this reason, it is useful to also define a second set of quantities we call generalized order parameters $m^\mu$ (denoted with an \emph{upper} index) that account for the fact that gene expression profiles of closely related cell fates are correlated \cite{kanter1987associative}. To define this generalized order parameter, first we define the matrix
\begin{align}
A_{\mu \nu}= \sum_{k=1}^N\xi_{\mu k} \xi_{\nu k},
\end{align}
which measures the pairwise similarity between different cell fates. The generalized order parameters can be defined in terms of the inverse of this matrix through the equation
\begin{align}
m^{\mu}(t)= \sum_{\nu=1}^p (A^{-1})^{\mu \nu} m_{\nu}(t)=\frac{1}{N} \sum_{j,\nu} (A^{-1})^{\mu \nu} \xi_{\nu j} x_j(t)
\label{def:mmu}
\end{align}
Unlike magnetizations, the generalized order parameters $m_\mu$ for two highly related cell fates are generically very different because of the presence of $A^{-1}$ in the expressions above.

The order parameters $m^\mu$ also serve as a natural coordinate system for cell fate space. This is because Eq.~\ref{def:mmu} has a geometric interpretation in terms of the linear projection of $x_i(t)$ onto the $p$-dimensional subspace spanned by the cell type vectors $\{ \xi_ {\mu j} \}_{\mu= 0,...,p-1}$ \cite{yampolskaya2023sctop}. In the appendix, we show any $x_i(t)$ can be written as
\begin{align}
x_i(t) = \sum_{\alpha=1}^p m^{\alpha}(t) \xi_{\alpha i} + x^{\perp}_i(t),
\label{Eq:xdecomp}
\end{align}
where $m^{\alpha}$ is the generalized order parameters for cell fate $\alpha$ and $x^{\perp}_i(t)$ is the part of gene expression vector that is perpendicular to p-dimensional subspace spanned by $\{ \xi_{\mu j} \}_{\mu=0,\ldots,p-1}$ \cite{yampolskaya2023sctop}. In other words, the first term represents the gene expression relevant to cell fate, while the second term contains all the gene expression information unrelated to cell type (such as the activity of housekeeping genes). Eq.~\ref{Eq:xdecomp} provides a natural way to transform between the $N$-dimensional gene expression space, $x_i(t)$, and the $p$-dimensional cell fate space, $\vec{m}(t)=(m^{1}(t), m^2(t),\ldots, m^p(t))$  (this corresponds to the top arrows in Figure \ref{cartoon}). 

To model developmental dynamics, we make use of the analogy between Modern Hopfield networks and developmental landscapes to formulate a dynamic update rule for gene expression of the form
\begin{align}
\tau{d x_i \over dt} &= \sum_{\mu=0}^{p-1}\xi_{\mu i}   \sigma^\mu( \beta m^{\mu}(t))-x_i(t),
 \label{Eq:xidynamics}
 \end{align}
with $\tau$ a time constant setting the speed of dynamics, $\sigma^\mu$ the soft-max function defined as 
 \begin{align}
  \sigma^\mu(\beta m^{\mu})= { e^{\beta m^\mu(t)}  \over \sum_\nu^p e^{\beta  m^\nu(t)}},
 \end{align}
and $\beta$ an inverse temperature parameter that controls the steepness of the non-linearity in the soft-max.  In the limit of zero temperature, this dynamical update rule ensures that the cell fates $\{ \xi_{\mu j} \}$ are fixed points. This can be seen by noting that when  $\beta \rightarrow \infty$, $\sigma^\mu(\beta \vec{m})=1$ for the direction $\mu=\gamma$ in cell fate space with the highest magnetization (i.e. $m^\gamma(t)=\max[\{m^\mu(t) \}]$) and zero otherwise, $\sigma^\mu(\beta \vec{m})=0$ if $\mu \neq \gamma$ (see SI section\ref{S-inverse temp}). 
 
We can also rewrite the dynamics in Eq.~\ref{Eq:xidynamics} entirely in cell fate space in terms of the $m^\mu$ by noting that
  \begin{align}
  \tau{dm^{\mu} \over dt} &= \sum_{j,\nu} A_{\mu \nu}^{-1} \xi_{\nu j} {dx_j \over dt}=  \sigma^\mu( \beta m^{\mu}(t) ) -m^\mu(t),
  \label{Eq:mdynamics}
 \end{align}
 where in the first equality we have used Eq.~\ref{def:mmu} and in the second equality we have used Eq.~\ref{Eq:xidynamics}. This expression shows that our differentiation dynamics do not change gene expression in directions $x^{\perp}_i(t)$ perpendicular to the p-dimensional subspace defined by the cell fates $\{\xi_{\mu i}\}$ (i.e., $\frac{dx_i^{\perp}}{dt} = 0$). For this reason, we will largely ignore $x^{\perp}_i(t)$ in what follows.

To make a connection with the Waddington landscape and as a prelude to incorporating signaling-induced cell fate transitions, it is helpful to rewrite Eq.~\ref{Eq:mdynamics} in terms of a ``potential function'' $V(t)$. In order to do so, we define a potential function of the form
\begin{align}
 V(t) = - \frac{1}{2} \sum_{\gamma=1}^p m^{\gamma}(t) m_{\gamma}(t).
\end{align}
and note that the dynamical Eq.~\ref{Eq:mdynamics} can be rewritten as 
\begin{align}
  \tau{dm^{\mu} \over dt} &= \sigma_{\mu} \left(- \beta {\partial V(t)\over \partial m_{\mu}(t)}\right) -m^\mu(t)
 \end{align}
This looks almost like damped gradient dynamics (with the $-m^{\mu}$ acting as a damping, friction-like term), with the key difference that the $p$-dimensional ``force'' vector $-{\partial V(t)\over \partial m_{\mu}(t)}$ is passed through a non-linear soft-max function.

In cell fate space, the potential $V$ is a inverted parabola centered at the origin. In the update rule, the state is pushed by a force caused by this potential, $ -\frac{\partial V}{\partial m_{\mu}}$, down along this inverted parabola. This potential has no minima, but the dynamics have a steady state because the non-linearity $\sigma_{\mu}$ bounds $m^{\mu}$ such that $|m^{\mu}|\leq 1$. The combined effect of the parabola and the bound on $m^{\mu}$ is that a state is pushed down the parabola and then hits the wall represented by the bound, reaching an attractor and remaining there because it can't be pushed farther along the parabola. This illustrates that we can think of this dynamics as a system pushed by a force defined by the gradient of a potential, but now rectified through a non-linearity.

In the dynamics defined by  Eqs~\ref{Eq:xidynamics} and \ref{Eq:mdynamics}, the p-cell fates of interest $\{ \xi_{\mu j} \}$ are always stable attractors. However, we know that during development cells can differentiate in response to external signals. During a cell fate transition,  cells can change their gene expression profiles from an initial attractor, for example a lung progenitor state, to another attractor corresponding to a more differentiated cell fate such as an alveolar cell. To incorporate this in the model, we add a signal-dependent potential $\tilde{V}$ to our potential of the form:
\begin{align*}
    V &= - \frac{1}{2} \sum_{\mu} m^{\mu}m_{\mu} + \tilde{V}(\vec{m}, f, k, ...).
\end{align*}
This signal-dependent potential $\tilde{V}$ is a function of the cell fate coordinates $\vec{m}$ as well as signaling parameters $f, k, ...$ which represent signals received by the cell. The parameters control the stability of attractors in the landscape. For example, an increase in parameter $f$ or $k$ might destabilize some attractors and stabilize others. The specific form of $\tilde{V}$ is constructed by considering different classes of decision-making. These classes, described by \citet{raju2023geometrical,rand2021geometry,camacho2021quantifying,saez2022statistically}, are broad categories that capture general features of cell fate transitions, such as the number of attractor states and the possible paths between them. The $\tilde{V}$ are universal functions constructed from normal forms of bifurcation and hence uniquely identify the decision making class associated with a cell fate transition.  Finally, since $\tilde{V}$ is a scalar function of the vector $\vec{m}$, it must be dependent on $\vec{m}$ in such a way that every raised index (e.g. $m^{\mu}$) is matched by a lowered index (e.g. $m_{\mu}$) so that the result is a scalar (see section \ref{S-invariant transform} for a discussion of invariant transforms and the interpretation of these covariant and contravariant vectors). With these additions, the full dynamics of our model are:
\begin{align}
 \tau {dx_i \over dt} &= \sum_{\mu}\xi_{\mu i} \sigma_{\mu} (-\beta \frac{\partial V(t)}{\partial m_{\mu}}) -x_i(t)\\
 &= \sum_{\mu}\xi_{\mu i}  \sigma^\mu(m^{\mu}(t) -\frac{\partial \tilde{V}(t)}{\partial m_{\mu}})-x_i(t),
\label{Eq:fulldynamics1}
\end{align}
or in cell fate space
\begin{align}
\tau {d m^\mu \over dt} &= \sigma^\mu(-\beta \frac{\partial V(t)}{\partial m_{\mu}}) -m^\mu(t).
 \label{Eq:fulldynamics2}
\end{align}
Note, that these dynamics no longer possess a Lyapunov function and the right hand side can no longer
be written as a gradient of a potential function due to the soft-max function $\sigma^\mu$.

Our mathematical construction is summarized in Figure \ref{cartoon}. The top row of Figure \ref{cartoon} shows the relationship between gene expression and cell fate coordinates. The cell type gene expression profiles $\xi_{\mu i}$ provide a way to translate between gene expression $x_i$ and cell fate coordinates $m^{\mu}$. The middle row of figure \ref{cartoon} illustrates how the dynamics between the spaces of gene expression and cell fate are related. As in the dynamics of a Hopfield model, the stored patterns -- or cell fates -- $\xi_{\mu i}$ provide a way to move between the spaces of gene expression and cell fate: $\frac{dx_i}{dt} = \sum_{\mu}\xi_{\mu i} \frac{d m^{\mu}}{dt}$. The dynamics $\frac{dm^{\mu}}{dt}$ involve the force exerted by the cell fate landscape $V$, which is controlled by the signaling parameters $\{f, k, ...\}$. These parameters define an abstract signaling space, and they represent chemical and mechanical cues that cause the cell to change fates. By featuring these three spaces -- gene expression, cell fate, and signaling -- our model is able to predict trajectories corresponding to landscapes of any decision-making class.

\subsection{Constructing potentials from elementary bifurcations}

\begin{figure}[p]
  \centering
  \includegraphics[width=0.4\textwidth]{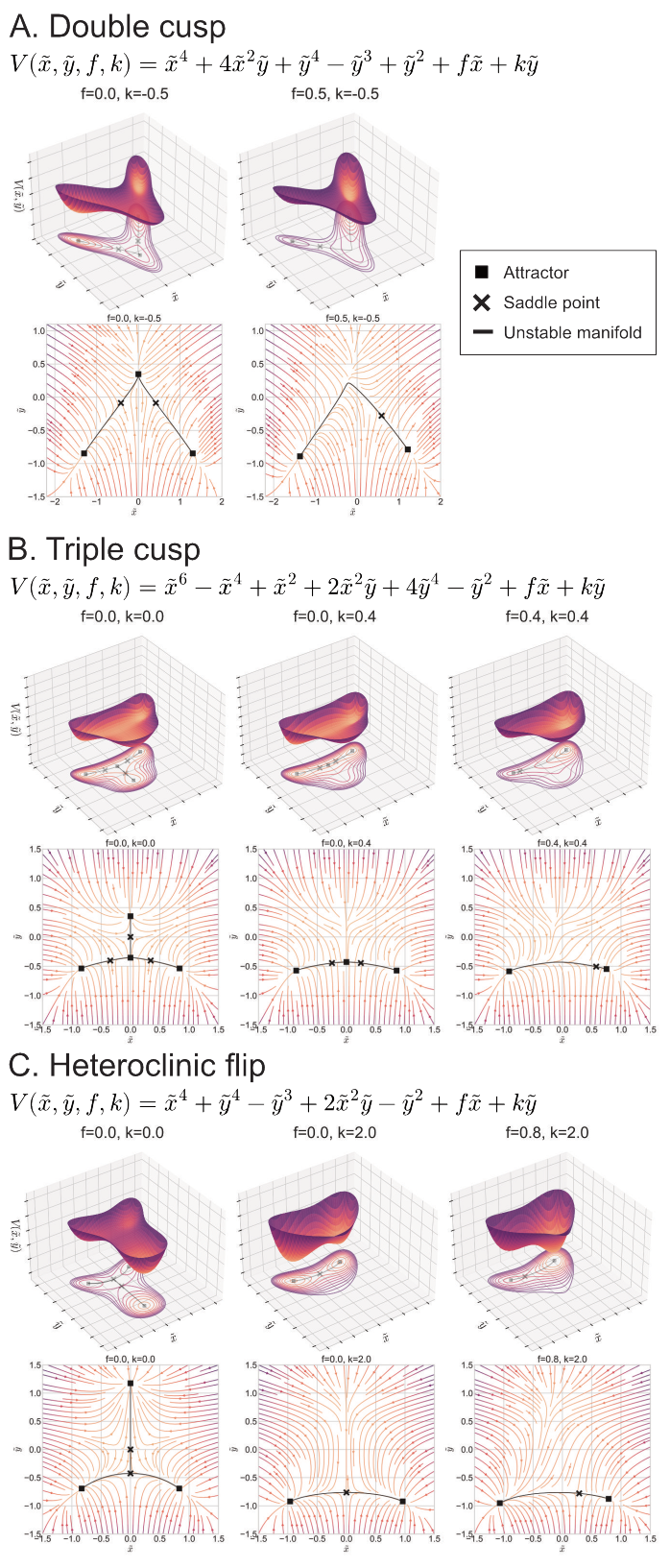}
  \caption{{\bf Three cell fate decision-making classes and representative landscapes.} Each class corresponds to a different way of connecting attractors and saddle points, shown via the potential $\tilde{V}$ on a two-dimensional landscape. Bifurcation parameters $f, k$ (varied left to right) represent signaling conditions that stabilize or destabilize attractors. Below each landscape, a graph shows the underlying decision structure. The attractor highest on the $\tilde{y}$-axis is taken as the initial cell type; squares are attractors, crosses are saddle points, and lines are unstable manifolds. {\bf A.} {\it Double cusp:} three attractors and two saddles, permitting only direct transitions from the initial attractor to either final state. Varying $f$ collides the middle attractor with a saddle, destroying both. {\bf B.} {\it Triple cusp:} four attractors and three saddles. The initial attractor leads to an intermediate attractor between the two final states; $k$ destabilizes the bottom attractor and $f$ destabilizes one of the top two. {\bf C.} {\it Heteroclinic flip:} three attractors and two saddles, connected so that a cell can be redirected to either fate given the right signals before committing. $f$ controls the stability of the bottom attractor; $k$ tilts the landscape toward one of the top attractors.}
  \label{bifurcation diagrams}
\end{figure}

The model described in the previous section combines Hopfield dynamics with a landscape. The question, then, is how to construct a landscape. Landscapes can be defined by the bifurcations they contain. A bifurcation occurs in a dynamical system when a parameter is varied and the stability of a fixed point is changed as a result. The number of parameters one varies to cause the bifurcation is called the codimension of the bifurcation. All local bifurcations of codimension less than or equal to 5 have been enumerated, and they are called elementary catastrophes \cite{thom2018structural,zeeman2006classification}. Generic forms for landscapes of these elementary catastrophes are called normal forms, and they take the form of polynomials. For example, one of the simplest bifurcations is the codimension-1 fold, or saddle node, bifurcation, which has the normal form $V(x) = x^2 + a$. $a$ is the bifurcation parameter. When $a<0$, there are two fixed points; when $a>0$, there are none. 

To construct landscapes for cell fate decisions, we follow the method described by \citet{rand2021geometry}. Their work sets the mathematical foundation for identifying universal features of cell fate transitions. We provide a conceptual overview of their method of constructing landscapes. Landscapes are categorized by decision-making classes defined by numbers of attractors, saddle points, and the paths between them. The structure of each class can be depicted as a decision graph showing the paths between attractors and saddle points (see Figure \ref{bifurcation diagrams}). Instead of using all of the elementary catastrophes, these decision-making classes are constructed by combining two simple kinds of bifurcations: folds and heteroclinic flips. These two bifurcations represent the basic ways one can alter decision-making classes: fold bifurcations create or destroy stable fixed points and saddle points, while heteroclinic flips change which stable fixed points are connected. They are sufficient for constructing any class that can be represented in the decision graph structure \cite{rand2021geometry}. Complicated higher-codimension bifurcations like the swallowtail and butterfly catastrophes are neglected because they would require fine tuning multiple parameters which is assumed to be evolutionary unlikely \cite{bialek2012biophysics}.

Formally, this argument relies on the properties of gradient-like Morse-Smale systems: dynamical systems that are structurally stable and in which every trajectory ends at a fixed point \cite{raju2023geometrical}. These properties match the picture of the Waddington landscape, where every cell attains its fate in a robust, controlled manner. Structural stability means that trajectories do not change drastically under small perturbations, except at bifurcations. As discussed above, small changes in bifurcation parameters generically cause significant changes in behavior, and a bifurcation causes the system to transition from one Morse-Smale system to another. The only bifurcations needed to connect similar Morse-Smale systems are the fold and the heteroclinic flip \cite{raju2023geometrical,newhouse1976there}.

The equations describing the landscapes for these classes are based on the normal forms of the elementary catastrophes. In other words, one takes the unfolding of a sufficiently high dimension and adjusts the parameters of terms to match the behavior of the decision graph. For example, the elliptic umbilic $\tilde{V} = a y^3 - b x^2 y + cx^2 + dy^2 + ky + fx$ serves as the starting point for writing out the landscapes described below (the corresponding equations are shown in Figure \ref{bifurcation diagrams}) \cite{rand2021geometry}. The landscapes below are found by modifying this normal form in two ways. The first modification is to add higher-order even terms ( $x^4$, $y^4$, $x^6$, $y^6$) so that $\tilde{V}\rightarrow \infty$ as $x\rightarrow\infty$  or $y\rightarrow\infty$. This puts the system in a potential well so that it doesn't escape to infinity. The second modification is to choose the coefficients of each higher-order term ($a, b, c, d$) to match the behavior of the chosen decision graph. After this adjustment, the only bifurcation parameters are the coefficients of the linear terms $f, k$. These represent the two parameters controlled by signals received by the cell. Their effect mathematically is to tilt the landscape along each of the axes. As shown in Figure \ref{bifurcation diagrams}B for the case of the triple cusp landscape, tilting along the $y$-axis corresponds to a change in $k$ that destabilizes the initial attractor basin. A change in $f$ tilts the landscapes along the $x$-axis which biases the decision towards one of the final two fates. 

The parameters $f$ and $k$ can be defined explicitly in terms of biological signals; for example, each could be a function of the Notch ligand produced by a cell's neighbors and received by its receptors (see the section on spatial patterning below). For most of the examples in this paper, however, we leave $f$ and $k$ as abstract parameters that determine the stability of attractors and control them manually in simulations, as described in Supplementary Information section \ref{S-signal dynamics}.

We now describe some specific cases of decision-making classes, characterized by their decision graphs showing attractors, saddle points, and the paths between them. Biologically interesting dynamics require at least three attractors, since this allows the cell to choose between two fates as it leaves its initial state. The two simplest such classes are the double cusp and the heteroclinic flip.

The double cusp is shown in Figure \ref{bifurcation diagrams}A. It consists of three attractors and two saddle points, with the saddle points separating the initial attractor from each of the final two. A cusp bifurcation can be constructed from two fold bifurcations meeting at a saddle point, and a double cusp is built from two such cusps, or equivalently four intersecting folds. Shifting the bifurcation parameter $f$ destabilizes the initial attractor and biases the cell toward one of the final two fates.

The heteroclinic flip is shown in Figure \ref{bifurcation diagrams}C. It also has three attractors and two saddle points, but here the saddle points are connected to one another. In this arrangement, the parameter $k$ controls the stability of the initial attractor independently of $f$, which biases the cell toward one of the two final fates.

A third, more complicated class is shown in Figure \ref{bifurcation diagrams}B. This class includes a fourth attractor in the middle of the three primary attractors. This corresponds to a transient attractor which is a mixed state of the final two attractors. When $f$ is varied, the cell moves from the initial attractor to this intermediate attractor, and $k$ destabilizes this intermediate state to push the cell to one of the final two states. We include this class because there is evidence to suggest that some cell fates have a bipotent progenitor state which is a mixture of the final fates \cite{yampolskaya2023sctop,weinreb2020lineage}.

\subsection{Identifying experimental signatures of decision-making classes}

\begin{figure}[p]
  \centering
  \includegraphics[width=0.9\textwidth]{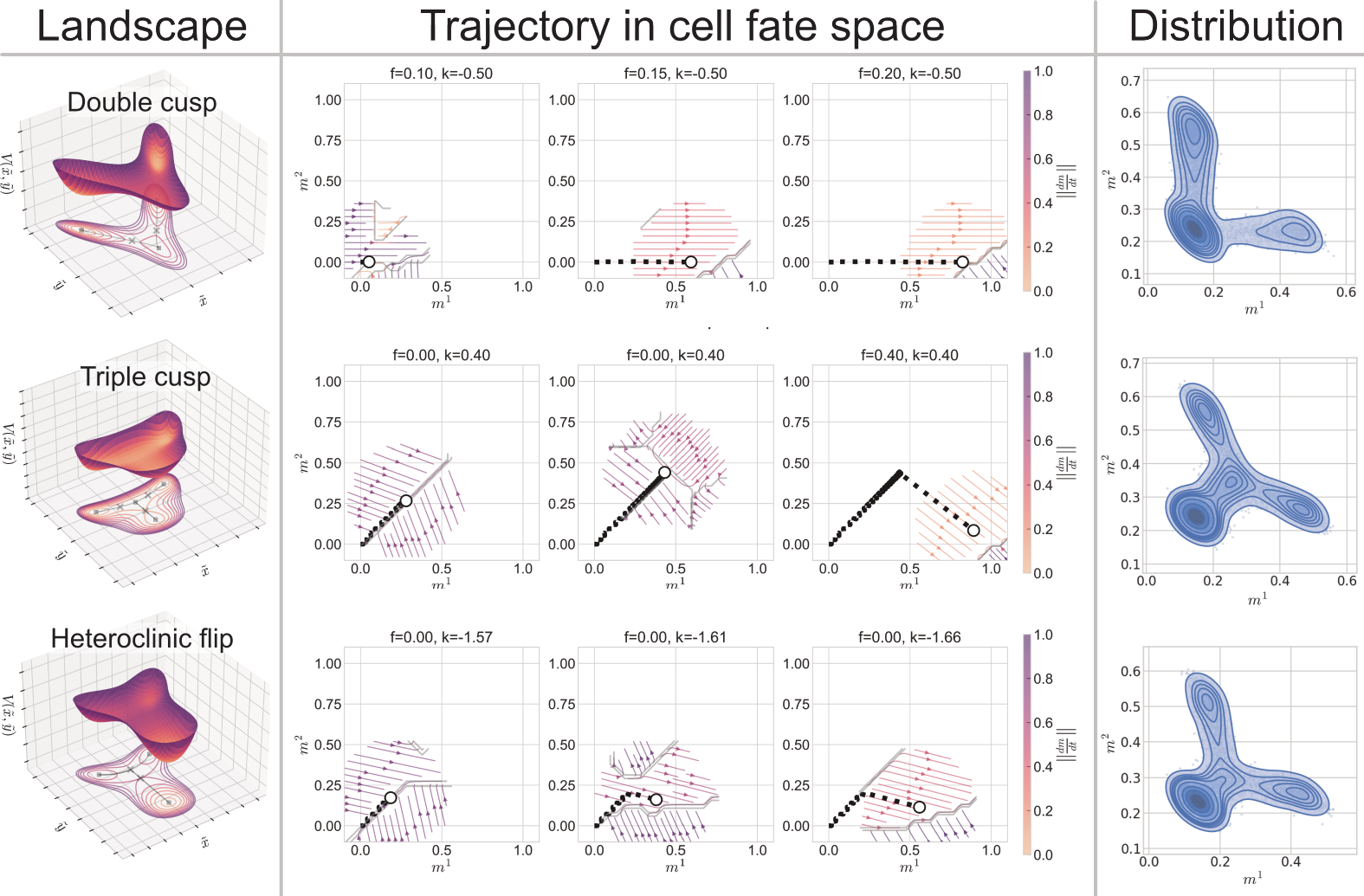}
  \caption{{\bf Three decision-making classes and typical trajectories in cell fate space.} The landscapes from Figure \ref{bifurcation diagrams} are shown on the left. In the middle, sample trajectories and vector plots corresponding to the dynamics of Equation \ref{Eq:fulldynamics2} are shown for cells starting in the initial state $m^0 = 1, m^1 = 0, m^2 = 0$ as the signaling parameters $f,k$ are varied. The white shows the position of the cell under each set of control parameters $f,k$; the dotted line shows the trajectory of where the cell has been; the streamlines illustrate the directions toward which the landscape bias cells (for clarity, only a small area around current position is shown ). Nullclines are shown in gray. On the right, contour plots showing the density of trajectories for many simulated cells, highlighting distribution differences across the three classes (see Figure \ref{S-fig:noisy classes} for scatter plots).}
  \label{signatures}
\end{figure}

Once we have constructed a landscape for a given class, we can simulate cell fate transitions using our Hopfield-like dynamics. This produces archetypal trajectories that highlight each class's unique features and demonstrate possible signatures of saddle points and intermediate attractors in time-series data. The middle column of Figure \ref{signatures} shows these trajectories along with the local vector field, which reveals the surrounding attractor basins and the stability of different regions. For example, the region of multilineage expression (where more than one cell fate coordinate is nonzero) is inaccessible in the double cusp class but accessible in the other two, so our model predicts different behavior for cells starting in this region depending on the class. The right column shows contour plots from many simulated trajectories, mimicking scRNA-seq data where cells are sampled from many trajectories and time points. Cells cluster as they approach stable regions, producing peaks around attractors, while steeper unstable regions contain fewer cells.

It is important to note that the topology of a class does not fully determine the trajectory; many trajectories are possible for each landscape, depending on the relative positioning of saddles and attractors. For demonstration, we show results for some ``typical'' (i.e.\ not pathological or extreme) topographies that emphasize the distinctions between classes. An exhaustive exploration of how trajectories imply or exclude bifurcation classes is beyond the scope of this work and is an important direction for the future.

Comparing these typical trajectories across landscapes reveals potential signatures for each decision-making class. In the double cusp, signals push cells directly to their final fates, producing straight-line trajectories. In the triple cusp, cells spend more time in the transiently stable intermediate state than elsewhere along the trajectory, producing a localized high-density cluster that serves as the defining feature of this class. The heteroclinic flip is identified by curved paths, which cells follow as they move away from the second saddle node; the unstable manifold between attractors makes this class especially sensitive to fate-specific signals, as seen in the contour plot where cells quickly leave the intermediate region. By searching for these signatures in data, one can directly compare real transitions to theoretical landscapes.

The cell fate coordinates $m^{\mu}$ can be measured in real systems by applying scTOP to scRNA-seq data \cite{yampolskaya2023sctop}. Time-series data is particularly valuable, since identifying features of dynamical systems often requires temporal information; for example, the time cells spend in different states distinguishes transient from stable attractors. In embryonic development, snapshots across time points reveal the path of cells as signaling and mechanical conditions vary. In this section, we look for signatures of decision-making classes in two contexts: \textit{in vitro} mouse hematopoiesis and the developing mouse lung.

\subsubsection{Cell fate decisions in hematopoiesis}

Blood requires frequent replenishment, so a wide range of blood cells are regularly created in bone marrow. These blood cells originate from hematopoietic stem and progenitor cells (HSPCs), which have the capacity to differentiate into many blood cell fates. They are responsible for maintaining stable populations of blood cells. While many different cell fate hierarchies have been proposed for hematopoiesis, lineage tracing data from \citet{weinreb2020lineage} suggests that HSPCs primed to become different cell types may exist on a continuum rather than passing through distinct progenitors for each mature fate. For example, some HSPCs which become monocytes take paths which are closer to dendritic cells while others take neutrophil-like paths. By comparing populations of HSPCs and the paths they take in cell fate space, our model can identify signs of landscapes associated with these possible scenarios.

Lineage tracing experiments in \citet{weinreb2020lineage} enable a precise analysis of progenitors and their eventual fates. HSPCs were extracted from the bone marrow of mice and tagged with DNA barcodes unique to each cell. These barcodes are retained when cells divide. Since scRNA-seq destroys the cell it samples, it's difficult to observe the gene expression of a particular cell over time. Lineage tracing by barcoding is one way to bypass this difficulty, since it allows mature cells to be matched to clonal progenitors via the barcode. In this experiment, after cells were barcoded and allowed to proliferate ex vivo, some cells were sampled, and some were replated and allowed to divide and differentiate further. This process was repeated twice, which led to samples being taken 2, 4, and 6 days after barcoding (see Figure \ref{signatures in data} A). This process allows the identification of clonal families: clones at day 2 who have sister cells at days 4 and/or 6. 

To analyze this data with scTOP, we followed the procedure described in \citet{yampolskaya2023sctop}. In brief, we constructed a reference basis $\xi$ by taking averages across particular populations of cells from day 2 and day 6. For each subpopulation, we averaged the gene expression profiles for 150 random cells, excluding cells that had sister clones on other days. From day 2, we sampled undifferentiated cells, while at day 6 we sampled mature cells with labels provided by the author. This resulted in a reference basis with seven types: progenitors (HSPCs), neutrophils, monocytes, megakaryocytes, mast cells, eosinophils and basophils. Although HSPCs may only be transiently stable, scTOP was able to accurately identify them with this reference basis, which validated treating HSPCs as effective (if temporary) attractors.

HSPCs have many possible fates, and landscapes that involve many attractors can be highly complex. Including landscapes of cells that differentiate into more than two fates would require further extension of this work into exploring the decision-making classes in higher dimensions. To compare this data with landscapes containing only three attractors, we limited our analysis to progenitors that are effectively bipotent. In other words, only clonal families which involved exactly two final fates were included. We projected the multi-day trajectories of these cells onto our reference basis to obtain the cell fate coordinates corresponding to progenitor and mature cell types. Figure \ref{signatures in data}C, E shows clonal families that result in basophil/neutrophil and monocyte/neutrophil populations. The basophil/neutrophil cells take straight paths to their eventual fates. Since these cells don't cross into the intermediate zone of multilineage expression, where more than one cell fate coordinate is nonzero, it is unlikely that a saddle point or intermediate attractor exists between these fates. These straight paths suggest that the underlying dynamics flow directly to the final fates without passing through a mixed state, which matches the predictions of the double cusp class.

In the case of differentiating monocyte/neutrophil cells, the trajectories towards final states are curved, which means the cells pass through multilineage-expressing states before fully specifying. This suggests there is a saddle point or an attractor between these final fates. Although it is difficult to tell with the imbalance of data, the paths appear to diverge quickly from the initial fate. This suggests an unstable manifold between the neutrophil and monocyte fates. It's possible that this is the signature of a heteroclinic flip class. Strictly speaking, it's also possible there is an intermediate attractor very close to the initial attractor, but since this intermediate attractor would be very transient it effectively acts like a saddle point. Additionally, the neutrophil-like path for monocytes noted in \citet{weinreb2020lineage} could be the result of a saddle point between monocytes and neutrophils.

Identifying signatures of these classes in hematopoiesis can aid in verifying the lineage tree of differentiation. If two fates exhibit a dual-cusp-like path, they are more distantly-related; in fact, it's possible to take two entirely unrelated cell fates and observe the dual cusp, since there will be no mixed states between them. On the other hand, classes like the heteroclinic flip and triple cusp indicate closer relations, which may indicate the cell fates are adjacent on the lineage tree. 

\subsubsection{Cell fate decisions in the developing lung}

With a variety of specialized cell types, the lung provides a rich system for studying cell fate decisions. Different transitions between these types have been established in homeostasis, injury, and transplant \cite{alysandratos2021epithelial}. In this section, we explore the case of alveolar maturation in murine development. In the mouse, cells in the foregut specify into lung epithelial cells around 9 days post-coitum (dpc) \cite{Hawkins_Rankin_Kotton_Zorn_2016}. The alveoli, where gas is exchanged between the organism and the environment, appear around 16.5-18.5 dpc. Eventually, these cells differentiate into the two fates necessary for proper alveolar function: alveolar type 1 (AT1) and alveolar type 2 (AT2) cells. The lung continues to remodel and develop even after birth as the newborn animal begins to breathe \cite{zepp2021genomic}.

\begin{figure}
  \centering
  \includegraphics[width=0.9\textwidth]{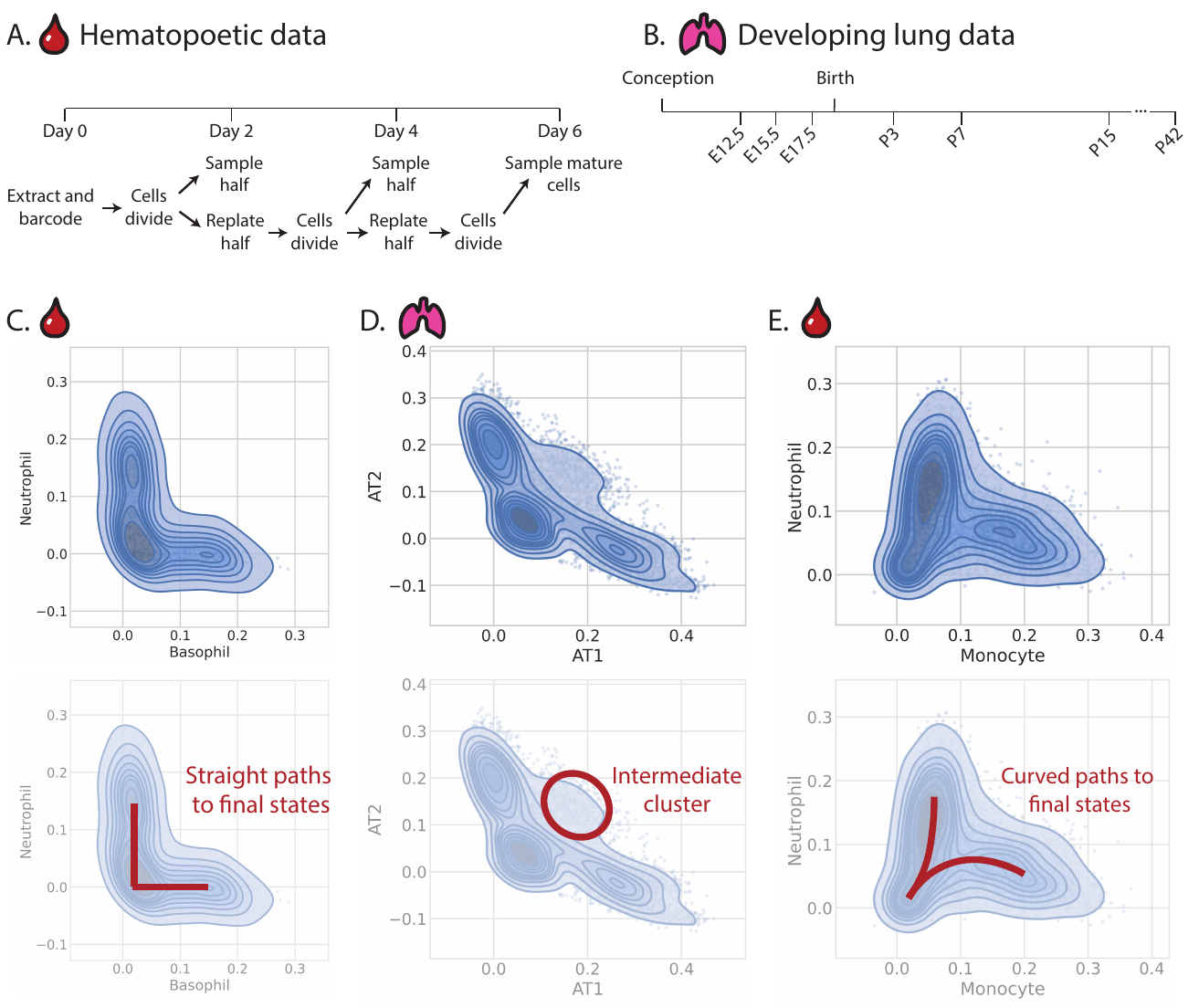}
  \caption{{\bf Experimental scRNA-seq data and corresponding landscape signatures.} {\bf A.} In \citet{weinreb2020lineage}, hematopoietic progenitor cells from mouse bone marrow were barcoded and cultured ex vivo; some were sampled by scRNA-seq on days 2 and 4, and mature blood cells were sampled on day 6. {\bf B.} \citet{zepp2021genomic} sampled mouse lungs from embryonic day 12.5 (E12.5) through postnatal day 42 (P42). {\bf C.} Basophil--neutrophil progenitors on basophil and neutrophil fate coordinates; the straight trajectories suggest a double cusp (top, right panel of Fig.~\ref{signatures}). {\bf D.} Developing lung cells on AT1 and AT2 fate coordinates; the intermediate cluster suggests a triple cusp (middle, right panel of Fig.~\ref{signatures}). {\bf E.} Monocyte--neutrophil progenitors on monocyte and neutrophil fate coordinates; the curved paths suggest a saddle point and hence a heteroclinic flip (bottom, right panel of Fig.~\ref{signatures}). Corresponding scatter plots colored by time point appear in Figure \ref{S-fig:data scatter}.}
  \label{signatures in data}
\end{figure}

\citet{zepp2021genomic} performed scRNA-seq on mouse lung samples from 12.5 dpc to 42 days after birth to investigate this complex process (as shown in Figure \ref{signatures in data} B). To track the differentiation of alveolar cells, we used a reference basis consisting of the combined Mouse Cell Atlas/Kotton laboratory basis provided in scTOP (the accuracy of which was validated in \citet{yampolskaya2023sctop}) as well as an additional reference cell type corresponding to 12.5 dpc epithelial progenitor cells from the \citet{zepp2021genomic} data \cite{han2018mapping,herriges_durable_2022}. For our analysis of the \citet{zepp2021genomic} data, we included all samples annotated by the authors as alveolar progenitors, AT1 cells, and AT2 cells. We projected these developing samples on the Mouse Cell Atlas/Kotton/epithelial reference basis to obtain cell fate coordinates spanning hundreds of cell types, and we found that only alveolar-related coordinates had non-negligible scores. The cell fate trajectories of developing alveolar cells across all time points are shown in Figure \ref{signatures in data} D. 

The main peaks in the data correspond to undifferentiated, AT1, and AT2 clusters. However, the differentiating cells also span a wide range of intermediate states. There are cells on direct paths to the mature fates as well as a small but non-negligible cluster right in the center of AT1/AT2 fates. These mixed-state cells suggest a transiently-stable attractor state. While it's possible this is a saddle point, our simulations of archetypal trajectories suggest that the vast majority of cells would drop off the unstable manifold and move towards one of the final fates with even a small amount of signal. The cluster exactly in the middle is easier to achieve in our model if we involve a transient attractor, as in the triple cusp class. Additionally, in some regimes of signaling space, either the triple cusp or heteroclinic flip can lead to direct paths to the final fates. Although these trajectories do not give conclusive evidence, compared to archetypal trajectories with simple topographies, they bear most resemblance to the triple cusp.

In our model, it is possible the mixed-state cluster is either an intermediate attractor or a saddle point, although the former requires less of a careful balance between signals in our simulations. Although the existence of multilineage expression is well-known, studying time series trajectories on cell fate coordinates and comparing them to simulations provides additional information about these mixed states beyond their mixed identities. With further study, we can learn about the dynamic properties of mixed states, such as stability and susceptibility to signals.  In the population that passes through the transient attractor state, cells move toward mature AT1 and AT2 fates shortly after birth. Given the timing of appearance and disappearance of this intermediate cluster, we hypothesize that the mixed state is an attractor that is stable until the mouse is born, and the remodeling associated with air-breathing destabilizes it post-birth.

\subsection{Decision-making classes and spatial patterns}\label{spatial}

\begin{figure}[p]
  \centering
  \includegraphics[width=0.6\textwidth]{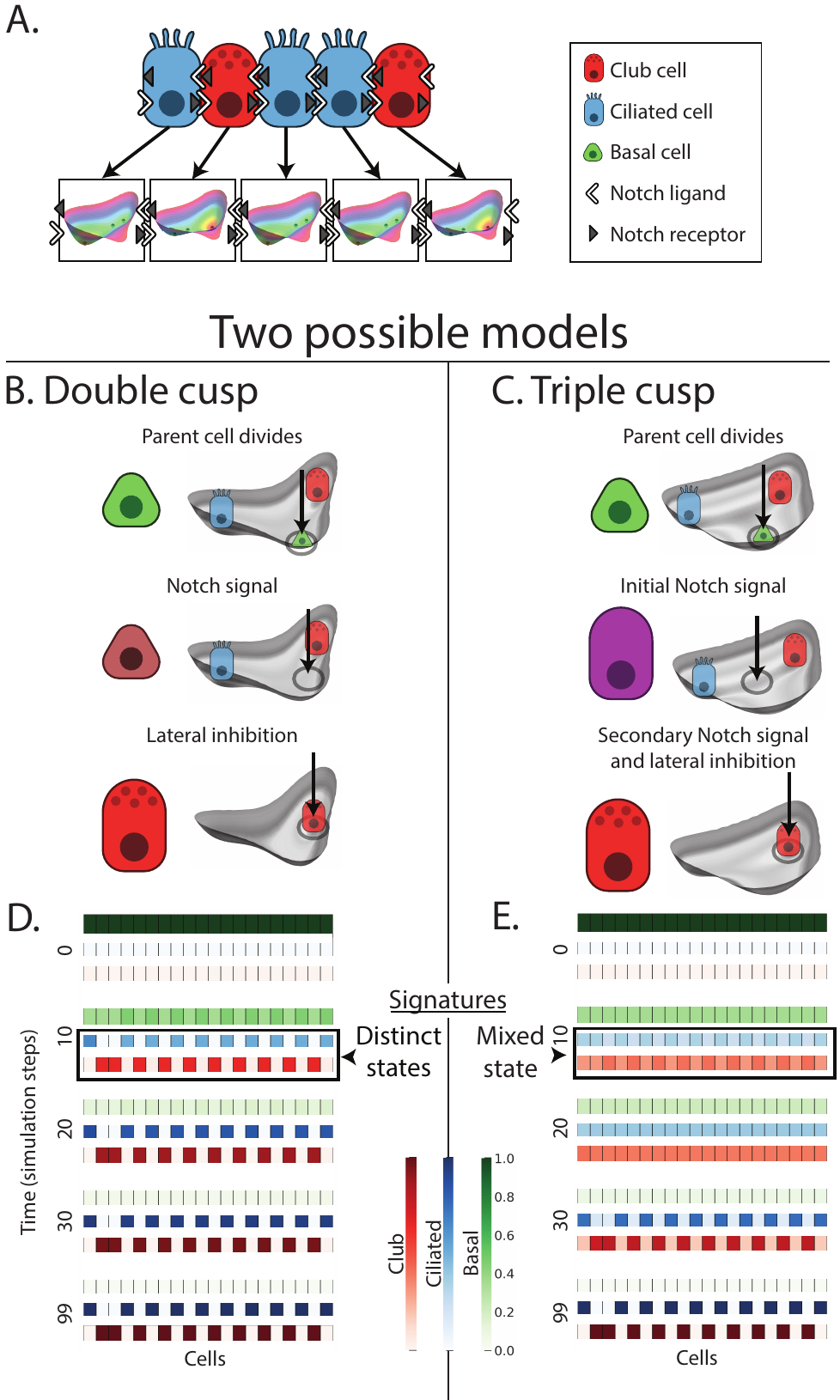}
  \caption{{\bf  Spatial patterning in the mouse airway post-injury.} {\bf A.} Intercellular signaling creates a pattern of club and ciliated fates. In our framework, each cell experiences its own landscape, tilted by signals from neighbors; lateral inhibition then drives alternating club and ciliated fates.{\bf B, C.}Two possible models of basal cell differentiation post-injury \citep{rock2011notch} corresponding to distinct decision-making classes: a double cusp class, where a single Notch signal initiates differentiation and lateral inhibition enforces the alternating pattern; and a triple cusp class with two signals, where the first sends cells to an intermediate state and the second, combined with lateral inhibition, produces the pattern. {\bf D, E.} Simulated trajectories of 20 cells at 5 time points reveal spatial signatures of each class. Rows show alignment with basal (top), ciliated (middle), and club (bottom) fates ($m_{basal}, m_{ciliated}, m_{club}$). {\bf D.} Double cusp: cells do not pass through an intermediate stage, instead increasing alignment only with their eventual fates. {\bf E.} Triple cusp: at step 10, cells transiently align with both club and ciliated fates, a mixed state that signals this class.}
  \label{spatial fig}
\end{figure}

Up to this point, we have considered the differentiation of individual cells, with external signals represented by the abstract parameters $f, k$ that we vary manually. In reality, many organs exhibit complex patterns of cell types. For example, in mouse and human lung airways, secretory cells produce mucus and anti-inflammatory proteins while ciliated cells use these secretions to clear the airways; a healthy mixture of both is essential for airway function.

The Notch signaling pathway is vital for creating this mixture of cell types \cite{whitsett2011notch}. When a cell receives Notch signaling from its neighbors, its own Notch expression is inhibited, a process known as lateral inhibition. In ciliated-secretory patterning, as in many other cell type patterns \cite{bruckner2024information}, luminal airway cells receiving low Notch signals become ciliated, while those receiving high Notch signals become secretory cells such as club cells, producing an alternating pattern. Remarkably, if the airway is injured and this layer is removed, the alternating pattern re-emerges as basal cells proliferate and differentiate \cite{rock2009basal,pardo2015injury}.

Because multicellularity is central to cell type pattern formation, we extended our model to include intercellular interactions. Combining lateral inhibition with our Hopfield landscape model recreates the alternating secretory-ciliated pattern seen in the airway. We simulated one-dimensional chains of cells, each experiencing its own landscape that tilts toward one fate or another according to signals received from neighbors (Figure \ref{spatial fig}A). Rather than controlling $f, k$ manually, we now make $f$ a function of the fates of each cell's neighbors. The attractor basins correspond to basal, ciliated, and club cells, and cells begin the simulation near the basal state. As they differentiate, they induce the alternate fate in their neighbors, producing a mixed club-ciliated pattern similar to those seen \textit{in vivo}. The endogenous airway shows less regularity than these simulations, likely due to inherent heterogeneity in how cells respond to signals \cite{goetz2024ability}; we assume uniform response for simplicity.

The landscape governing this process depends on how Notch signals guide basal cell differentiation. \citet{rock2011notch} identifies two preferred models. In the first, a single Notch signal initiates differentiation: after the parent basal cell divides, the daughter cells receive Notch signals that push them out of the basal basin and into either the ciliated or club fate. This corresponds to a double cusp landscape (Figure \ref{spatial fig}B), where the alternating pattern emerges from lateral Notch inhibition tilting each cell's landscape opposite to that of its neighbors.

In the second model, two Notch signals are required: the first converts the basal cell into an early progenitor, and the second commits this progenitor to a club or ciliated fate. This corresponds to a triple cusp (Figure \ref{spatial fig}C). The initial signal destabilizes the basal fate and pushes the cell into an intermediate attractor resembling a mixture of basal, club, and ciliated fates; a second signal destabilizes this intermediate attractor, and lateral inhibition tilts the landscape toward either the ciliated or club fate.

Both models produce the same final pattern, but the cell fate coordinates $m_{\mu}$ reveal differences during differentiation. In the double cusp, cells appear to commit to their eventual lineages early, showing heterogeneity immediately after the basal state is destabilized. In the triple cusp, cells remain homogeneous in a mixed state until the second Notch signal arrives. The presence of a distinct progenitor population can therefore be identified by checking how directly cells follow a path to their eventual fates. The two landscapes also imply different windows of sensitivity: the double cusp commits cells to a fate earlier, while the triple cusp can be biased toward either fate later in the process.

This example illustrates how landscape models connect to spatial patterning, though direct contact with experiments remains difficult. In addition to time-series signatures of the kind shown in Figure \ref{signatures}, spatial cell-type patterning offers another route to identifying classes. Spatial transcriptomics could provide cell fate information for neighboring cells, allowing direct observation of their spatial distributions.

\section{Discussion}

Dynamical systems theory provides a rich set of tools to capture the phenomenology of cell fate transitions with minimal parameters \cite{rand2021geometry,raju2023geometrical,saez2022dynamical}. Instead of requiring exact information on how each gene affects every other gene, the geometric picture shifts the focus to the most general features of transitions: attractors, saddle points, and flows. With this abstraction, one can define classes of landscapes with the aim of finding general principles underlying cell fate transitions. In this paper, we introduced a novel Hopfield-inspired construction for mapping between classes of landscapes, cell fate trajectories, and gene expression space. This mapping predicts unique trajectory features for different features, such as intermediate saddle points and transient attractors. Based on the discussed classes of decision-making and examples of corresponding topographies, we hypothesized signatures of archetypal trajectories for each class. Because our theory is rooted in the physics of Hopfield networks, we can use the corresponding order parameters as measures of cell fate space, providing common axes on which to compare simulations and real experiments. We analyzed hematopoietic and developing lung data and found possible candidates for each of the three classes. Basophil-neutrophil differentiation involved direct paths to final states, which follows the predictions of a basic topography corresponding to the double cusp. In neutrophil-monocyte differentiation, curves in the trajectories suggested an unstable manifold between the final fates, which matches the properties of a typical heteroclinic flip. Additionally, alveolar cell specification involved a mixed AT1/AT2 state that points to the possible existence of a transient attractor that is destabilized by the start of post-birth breathing. 

This paper centered on introducing our mapping between classes of low-dimensional landscapes and high-dimensional gene expression, and our hypothesis that trajectories in cell fate coordinates point to underlying features such as saddle points and attractors. We demonstrated the predictions of our model based on basic topographies showing features of these classes. However, we did not perform an exhaustive analysis of the precise implications of each class; mathematically, the topology of each class does not fully specify the topography or resulting trajectory. It is technically possible, for example, for the saddle points of the double cusp to be positioned anywhere between the initial and final fates. We restricted our discussion only to the simplest topographies for the sake of finding typical trajectories highlighting differences between classes. To prove our method of finding signatures in cell fate trajectories more rigorously, we proceed with a discussion of experimental predictions and ways to validate or falsify our hypothesis.

One concrete prediction resulting from our model is the case of cell fate patterning by lateral inhibition. As discussed in Section \ref{spatial}, signaling that inhibits a cell's neighbors' from adopting that cell's fate is a common way of achieving spatial patterns of fates. It is a simply mechanism for achieving an even distribution of, for example, ciliated cells or fly bristles.  We predict that the topography of the underlying landscape will determine the cell fate trajectories of cells undergoing this kind of development. To test this prediction, sampling the lung airway with small time steps over the course of differentiation post-injury and applying scRNA-seq to the tissues would reveal the true trajectory, which can subsequently be compared to the proposed signatures in Figure \ref{spatial fig}.

Another way to probe the underlying landscape is with careful \textit{in vitro} experiments involving the extraction of endogenous cells and application of signaling factors. Taking scRNA-seq data at time points throughout the process would allow for more accurate quantitative landscape-based modeling of these transitions by directly relating the effects of signals to displacements in cell fate space. The application of either chemical gradients or mechanical forces to cell populations that are repeatedly sampled would allow our models to predict the effects of these changes on measurable axes. This can also test the stability of states to identify whether they are saddle points, which are sensitive to perturbations, or attractors, which are fixed under the right conditions. One possible such experiment could involve a test of the triple cusp landscape with alveolar cells. By culturing the cells in the mixed AT1/AT2 state in the time point just before air breathing, one can test the hypothesis of a transient attractor that is destabilized post-birth. If the cultured cells do not differentiate past the mixed state, this suggests the presence of an attractor instead of a saddle point.

It is also possible to perform a more comprehensive validation of this idea through large-scale analyses of existing scRNA-seq atlases. Because our method can be applied to any scRNA-seq data with annotated cell types, it is easily scalable. With the increasing prevalance of atlases that contain many developmental time points, it is possible to systematically map out the cell fate trajectories of all developmental transitions for well-studied species \cite{wang2023construction,qiu2024single}. Then, one can seek the signatures laid out in this paper across all these transitions and explore the prevalence of different classes. With such a survey, one could probe whether it's possible to connect different signaling pathways, such as Notch and Wnt, to particular classes of landscapes. Our method of identifying signatures can be combined with existing statistical methods to find the most probable landscapes \cite{mochulska2025generative,cislo2025reconstructing}. Additionally, one could further test the existence of intermediate saddle points or attractors by identifying where potential heteroclinic flip or triple cusp trajectories appear in these atlases and targeting those cell types for further exploration \textit{in vitro}. Different classes predict different sensitivities to signals, so applying varying levels of signals can be another test of the landscape. The heteroclinic flip is the most sensitive to lineage-defining signals, since a small perturbation can push a cell towards one fate or the other. On the other hand, the typical topography for a double cusp would suggest that it is very difficult to push a cell into multilineage expression. 

Related work falls into two categories: models that identify classes of cell fate landscapes, and models that connect gene expression to cell fate via Hopfield networks. Our model unites these by using associative memory as the map from modular landscapes to dynamics, with Hopfield-inspired coordinates providing a universally applicable phase space. To our knowledge, no previous model links transcriptome-wide gene expression, classes of landscapes, and signals together, and this bridge yields generalizable insights into the data.
Our work complements existing models of decision topologies by predicting class-specific basins of attraction and regions of stability in a measurable phase space. The key ingredient is our use of continuous cell fate coordinates based on transcriptome-wide cell identity. In other works, such as \citet{saez2022statistically,mochulska2025generative}, cell identity is treated as categorical: cells are assigned to discrete fates by marker genes, and landscapes are fit to these proportions over time. This discards information about mixed-expression states that our continuous coordinates retain. \citet{cislo2025reconstructing} also use continuous coordinates, but only after curating a small set of relevant genes for each transition. Our method requires no gene selection and only minimal curation (cell type annotation, often already provided), making it straightforward to scale to hundreds of cell types \cite{yampolskaya2023sctop}. It also allows the same landscape to produce the same trajectory across different cell fate transitions, organs, and potentially species \cite{souza2026parameter}.
Several recent works draw parallels between associative memory and cell fate specification \cite{lang2014epigenetic,fard2016not,pusuluri2017cellular,guo2017hopland,ikonomou2020vivo,yampolskaya2023sctop,smart2023emergent,karin2024enhancernet,boukacem2024waddington,grishechkin2025mathematical}. Our construction is novel in incorporating bifurcation theory. In a standard Hopfield network, attractor stability is controlled only by temperature, a coarse-grained parameter \cite{grishechkin2025mathematical}; our model permits precise control over individual attractors and the modular addition of saddle points \cite{yampolskaya2023controlling}, which is necessary for modeling signal-driven differentiation. This also changes how multilineage expression is interpreted. In \citet{grishechkin2025mathematical}, multilineage expression appears as cells pass through saddle points during annealing. Our framework reveals additional possibilities: it can signal a saddle point, an intermediate attractor, or a connected sequence of such components, with paths ranging from the curved trajectories of monocyte-neutrophil progenitors to the 50/50 mixed states of AT1/AT2 cells.

Several extensions are natural. Statistical methods could be combined with our framework by fitting landscapes on our cell fate coordinates and assigning class probabilities. Theoretical work on bifurcation theory could extend the classification beyond binary decisions to multi-fate transitions. Finally, incorporating more realistic models of signaling would tie the framework more closely to experiments and enable sharper predictions.

\section{Methods}

\subsection{Defining cell fate coordinates}\label{cell fate coordinates}

 As previously described in \citet{yampolskaya2023sctop}, generalized Hopfield order parameters (called single-cell Type Order Parameters (scTOP)) can be applied to gene expression data to define cell fate space. In the case of a single cell, let $x_i$ denote the cell's expression of gene $i$. These are continuous values because, in scTOP, the RNA counts are converted to z-scores reflecting the rank-ordering of genes within a cell (e.g a gene $i$ at the 50 percentile is assigned $x_i=0$, a gene j at 84.6 percentile an expression level of $z=1$, etc). Let $\xi_{\mu i}$ be the expression of gene $i$ in cell type $\mu$.  With scTOP, $\xi$ is a list of gene expression profiles of known cell types. These are derived from scRNA-seq atlases such as the Mouse Cell Atlas and Tabula Sapiens by averaging across populations of cells, where each population corresponds to a different cell type \cite{han2018mapping, jones376tabula}. Then, as introduced by \citet{kanter1987associative}, the generalized Hopfield order parameters are defined as follows:
\begin{align*}
    m^{\mu} &= \sum_{\nu j} (A^{-1})^{\mu \nu} \xi_{\nu j} x_j \\
    A_{\mu \nu} &= \sum_{k}\xi_{\mu k} \xi_{\nu k}
\end{align*}
where $m^{\mu}$ is the order parameter which measures alignment with the $\mu$th cell type, and $A$ is the matrix of correlations between each cell type. 

The order parameters $m^{\mu}$ have a natural interpretation as a set of coordinates because they are equivalent to a projection onto the non-orthogonal subspace spanned by known cell types. The cell types $\xi$ form a subspace of gene expression space. When a cell's RNA counts are measured through scRNA-seq, this represents another vector in gene expression space. By projecting this vector onto the cell type subspace, we can describe it with a new set of coordinates, where each axis measures alignment with a cell type. In other words, $P_{ij} = \sum_{\nu \mu j}\xi_{\mu i} (A^{-1})^{\mu \nu} \xi_{\nu j}$ is a projection matrix that transforms vectors from gene expression space (denoted by $j$) to cell type space (denoted by $\mu$).  We can rewrite the original vector describing the gene expression of the cell as a decomposition onto this subspace:
\begin{align*}
    x_i &= \sum_{\mu} m^{\mu}\xi_{\mu i} + x_i^{\perp},
\end{align*}
with
\begin{align*}
    \sum_j P_{ij} x_j = \sum_{\mu} \left(\sum_{\nu \mu j} (A^{-1})^{\mu \nu}\xi_{\nu j} \right)\xi_{\mu i}  =  \sum_{\mu} m^{\mu}\xi_{\mu i},
\end{align*}
and
\begin{align*}
  x_i^{\perp}= x_i - \sum_j P_{ij} x_j.
\end{align*}

In the equation above, one can see that $m^{\mu}$ are the coefficients of decomposition onto the $\xi$ vectors, and $x_i^{\perp}$ is the component of $x_i$ that is perpendicular to the cell type subspace. This perpendicular component corresponds to gene expression information about processes that are not indicative of cell type, such as the expression of housekeeping genes. It corresponds to the information that is lost when going from the higher-dimensional space of gene expression to the lower-dimensional space of cell fate. In other words, $\sum_j P_{ij} x^{\perp}_j = 0$. This follows from the definition above by noting that for a projector $P^2=P$. In the classic Hopfield network, the stored patterns $\xi_{\mu i}$ are orthogonal, in which case $A_{\mu \nu} = \delta_{\mu \nu}$ and $m^{\mu} = \frac{1}{N} \sum_i\xi_{\mu i} x_i$. 

Since these generalized order parameters make use of a change of basis (from gene expression space to the basis of cell fate vectors), we use Einstein summation notation. The covariant and contravariant forms of the order parameters are distinguished by upper and lower indices. In this notation, we have a contravariant vector $m^{\mu}$, a covariant vector $m_{\mu}$, and a metric tensor $g_{\mu \nu}$ that is dependent on the correlations between patterns.  

We define the metric tensor using the product of our new basis vectors, $g_{\mu \nu} = A_{\mu \nu} = \xi_{\mu}\cdot \xi_{\nu}$, with the inverse of the metric tensor defined with upper indices: $g^{\mu \nu} = g_{\mu \nu}^{-1}$. For the covariant form (with a lowered index), we define $m_{\nu} = \frac{1}{N} \sum_{i} \xi_{\nu i} x_i$. For the contravariant form (with a raised index), we define:

\begin{align*}
    m^{\mu} &= \frac{1}{N} \sum_{i \nu} g^{\mu \nu} \xi_{\nu i} x_i \\
    &= \sum_{\nu} g^{\mu \nu} m_{\nu}
\end{align*}

\subsection{Choice of landscapes}
With this formulation, it is possible to simulate dynamics associated with any cell fate landscape. We consider landscapes associated with decision-making classes containing three primary attractors: a progenitor state with two possible mature fates. For simplicity and ease of visualization, this paper considers two-dimensional landscapes, $V(\tilde{x}, \tilde{y})$. However, in this model, three primary attractors corresponds to a three-dimensional cell fate space, since each dimension is associated with a cell fate. To align the attractors of a two-dimensional landscape with three-dimensional cell fate space, it is necessary to rotate the plane of the potential to align with the attractors. The details of transforming from $(\tilde{x}, \tilde{y})$ space to $(m^0, m^1, m^2)$ space are explained in \ref{S-coordinate transform}.

The gradient of the landscape $(\frac{dV}{d\tilde{x}}, \frac{dV}{d\tilde{y}})$ is a function of the signaling parameters $(k,f)$, which control the bifurcations.  The values for the signaling parameters $(k,f)$ were varied as follows. For the landscape with a triple cusp, the system starts with $k=0, f=0$. The signaling parameter $k$ is increased from 0 to 0.3, destabilizing the attractor in which the system started. Then, $f$ is changed from 0 to 0.3, causing the intermediate attractor to disappear and tilting the landscape towards one of the final fates. In the landscape with a double cusp bifurcation, the parameters start at $k=0.15, f=0$.  Parameter $k$ is kept at 0.15, and $f$ is changed from 0 to 0.1. This destabilizes the initial attractor and pushes the cell towards one of the final cell fates. The landscape with a heteroclinic flip bifurcation starts with $k=0.5, f=0$. Signal $k$ is shifted from 0.5 to 2 to destabilize the initial attractor, and before that bifurcation is fully complete, $f$ is changed from 0 to 0.5 to tilt the landscape towards the cell fate on the right-hand side. \ref{S-coordinate transform}.

\subsection{Simulation Details} 
To simulate Figure \ref{signatures}, we used scRNA-seq samples of alveolar type 1 and type 2 cells from \citet{herriges2023durable} and early epithelial samples from \citet{negretti2021single} to define the three cell fates (two mature fates and one progenitor state) for each of the three landscapes.

The initial and final values for the signaling parameters $(k,f)$ are specified for each simulation, along with the start and stop times for each signal. The signaling parameters are gradually changed from the initial to final values over the course of this defined signaling window (see SI section \ref{S-signal dynamics} for more details). It uses this information, along with the attractor states $\xi_{\mu i}$, to calculate the update step:
\begin{align*}
x_i(t+1) &= x_i(t) + \frac{dx_i}{dt}\\
\tau \frac{dx_i}{dt} &=  \sum_{\mu}\xi_{\mu i}  \sigma^\mu(m^{\mu}(t) -\frac{\partial V}{\partial \tilde{x}} \frac{\partial \tilde{x}}{\partial m_{\mu}} - \frac{\partial V}{\partial \tilde{y}} \frac{\partial \tilde{y}}{\partial m_{\mu}})-x_i(t)
\end{align*}
The derivatives $\frac{\partial \tilde{x}}{\partial m_{\mu}}, \frac{\partial \tilde{y}}{\partial m_{\mu}}$ come from applying the chain rule, and their specific forms are derived in SI section \ref{S-coordinate transform}. For each choice of landscape $V$, the simulated cell begins in the progenitor cell fate. Then, the signaling parameters $f, k$ are changed to destabilize the initial attractor state. Due to this bifurcation, the cell is pushed from its initial state. The parameter $f$ tilts the landscape towards one of the terminal fates, and the cell ends in the corresponding attractor basin. Please see Supporting Information for details.

\section{Data availability}
scTOP is available as a package on the Python Package Index (https://pypi.org/project/scTOP/) and the code is accessible on Github (https://github.com/Emergent-Behaviors-in-Biology/scTOP). The scRNA-seq data used in this paper come from a variety of sources. The alveolar lung data used for simulations is available on NCBI's Gene Expression Omnibus under accession code GSM6046035, as well as the Kotton Lab's Bioinformatics Portal (http://www.kottonlab.com)~\citep{herriges2023durable}. The early lung epithelial sample used as the simulated progenitor state is available under accession code GSE165063 \cite{negretti2021single}. The embryonic mouse lung data taken from E12.5 to P42 is available under accession code GSE149563~\citep{zepp2021genomic}. The hematopoietic lineage tracing data is available under accession code GSE140802, and the metadata can be found on Github (https://github.com/AllonKleinLab/paper-data/tree/master/Lineage\_tracing\_on\_transcriptional\_landscapes\_links\_state\_to\_fate\_during\_differentiation)~\citep{weinreb2020lineage}. 

\section{Code availability}
The Python and Mathematica notebooks used for the simulations and figures in this paper are available on Github (https://github.com/Emergent-Behaviors-in-Biology/Hopfield-landscapes).

\section*{Acknowledgments}
We acknowledge useful discussions with Jason Rocks, Michael Herriges, and members of the Kotton Lab and Mehta group. The work was funded by grants from the Boston University Kilachand Multicellular Design Program, Chan-Zuckerberg Investigator grant to PM, and NIH NIGMS 1R35GM119461 to PM.

\bibliographystyle{apsrev4-2}    
\bibliography{export-data}       

\end{document}


\section{Supplemental Information}

\tableofcontents 

\subsection{Mathematical details}

\subsubsection{Choice of potential} \label{choice of potential}
In choosing possible landscapes, we only considered landscapes containing three attractors. As explained in \citet{rand2021geometry}, two of the simplest bifurcations with three attractors are the binary cusp and the heteroclinic flip. However, in the case of AT1/AT2 differentiation, we noticed that there appeared to be a transient attractor. To account for this bifurcation, we also included the triple cusp. Bifurcation diagrams for these three bifurcations are shown in Figure \ref{M-bifurcation diagrams}.

Although these three types of bifurcations can occur in high dimensions, locally they are equivalent to two-dimensional normal forms; in other words, local models for bifurcations are universal, so we expect a two-dimensional landscape to capture the important qualities of a binary decision \cite{rand2021geometry}. For the binary cusp and heteroclinic flip, we use the same equations as \citet{saez2022statistically}. We found the equation for the triple cusp by introducing a higher-order term, $\tilde{x}^6$, and modifying the coefficients of the lower-order terms to produce behavior matching the decision graph. The equations for all three landscapes are shown in Fig. \ref{M-bifurcation diagrams}:
\begin{align*}
V_{\text{double cusp}}(\tilde{x}, \tilde{y}, f, k) &= \tilde{x}^4 + \tilde{y}^4 - \tilde{y}^3 + 4\tilde{x}^2\tilde{y}  +\tilde{y}^2 + f\tilde{x} + k\tilde{y} \\
V_{\text{triple cusp}}(\tilde{x}, \tilde{y}, f, k)  &= \tilde{x}^6 - \tilde{x}^4 + \tilde{x}^2 + 2\tilde{x}^2\tilde{y} + 4\tilde{y}^4 - \tilde{y}^2 + f\tilde{x} + k\tilde{y} \\
V_{\text{heteroclinic flip}}(\tilde{x}, \tilde{y}, f, k) &=  \tilde{x}^4 + \tilde{y}^4 - \tilde{y}^3 + 2\tilde{x}^2\tilde{y} - \tilde{y}^2  + f\tilde{x} + k\tilde{y}
\end{align*}

These landscapes are functions of two variables, $\tilde{x}, \tilde{y}$, and the way we translate between these variables and cell fate coordinates is described in Section \ref{coordinate transform}

\begin{figure}
  \centering
  \includegraphics[width=0.7\textwidth]{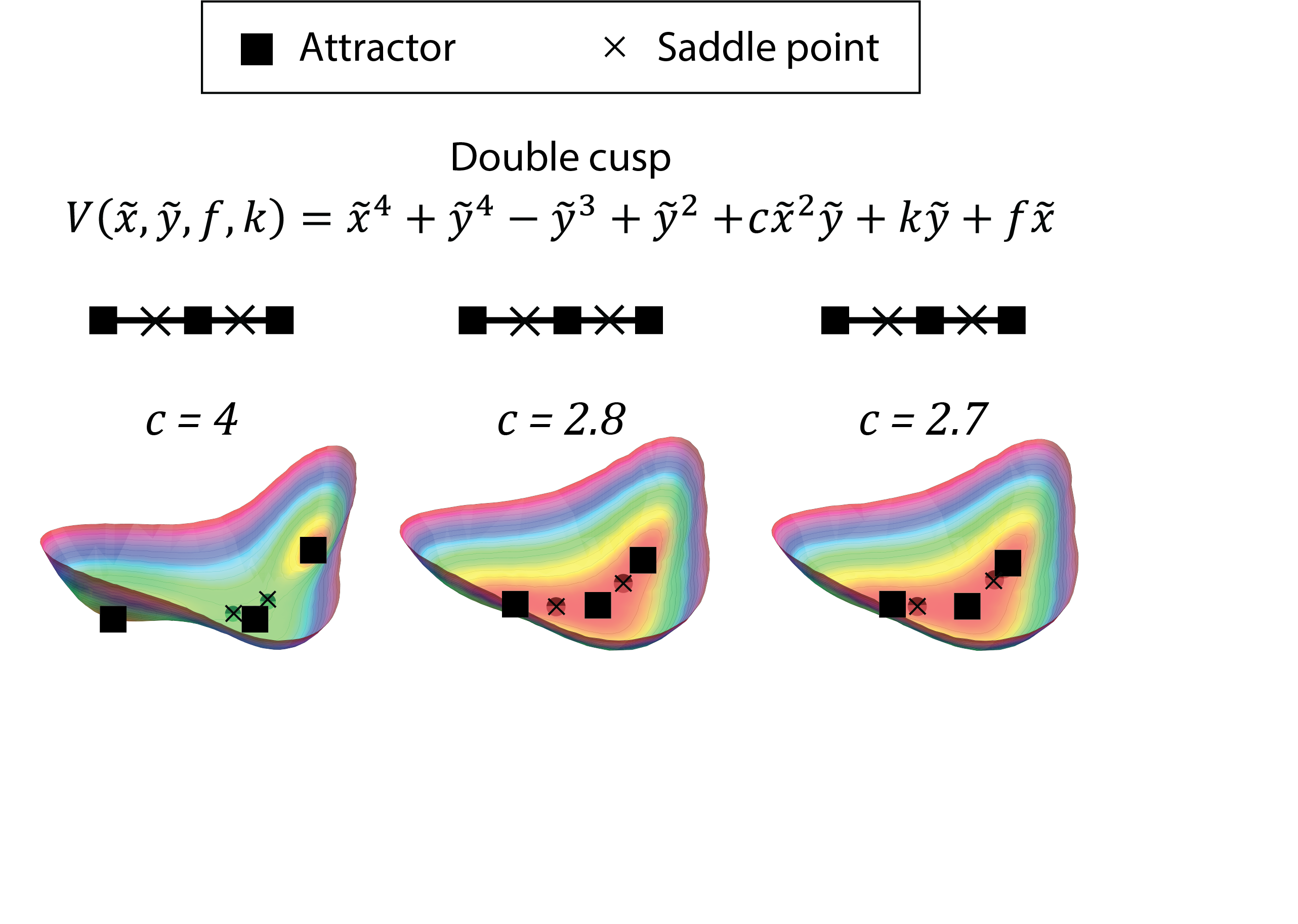}
  \caption{In the normal form of the double cusp, changing parameter $c$ varies whether the saddle nodes are closer to the initial attractor basin or the final basins. Note that the bifurcation diagram stays the same, since the diagram only conveys a topological description of the bifurcation and not the details of position.}
  \label{param change}
\end{figure}

Small changes to the coefficients of these forms modifies the positions of attractors and saddle nodes, which can change the shape of the landscape. For example, changing the coefficient of the $\tilde{x}^2 \tilde{y}$ term in the double cusp shifts the positions of the saddle nodes, as shown in Figure \ref{param change}. Varying $c$ slightly in this way doesn't change the bifurcation, but it will change the path a ball would take in this landscape and the necessary push needed for the ball to escape the first basin. In terms of cell fate, this corresponds to the strength of signal needed to transition from the initial fate to one of the final fates. Additionally, one can imagine an asymmetry between fates, where less signal is needed to transition to one of the final fates relative to the other. The class of bifurcation does not constrain the possible relative positions of attractor points and saddle nodes.

Large changes to the coefficients results in different bifurcation diagrams. As shown in Figure \ref{M-bifurcation diagrams}, there are regions of parameter space where all three landscapes have the same bifurcation diagrams.

\begin{figure}
  \centering
  \includegraphics[width=0.6\textwidth]{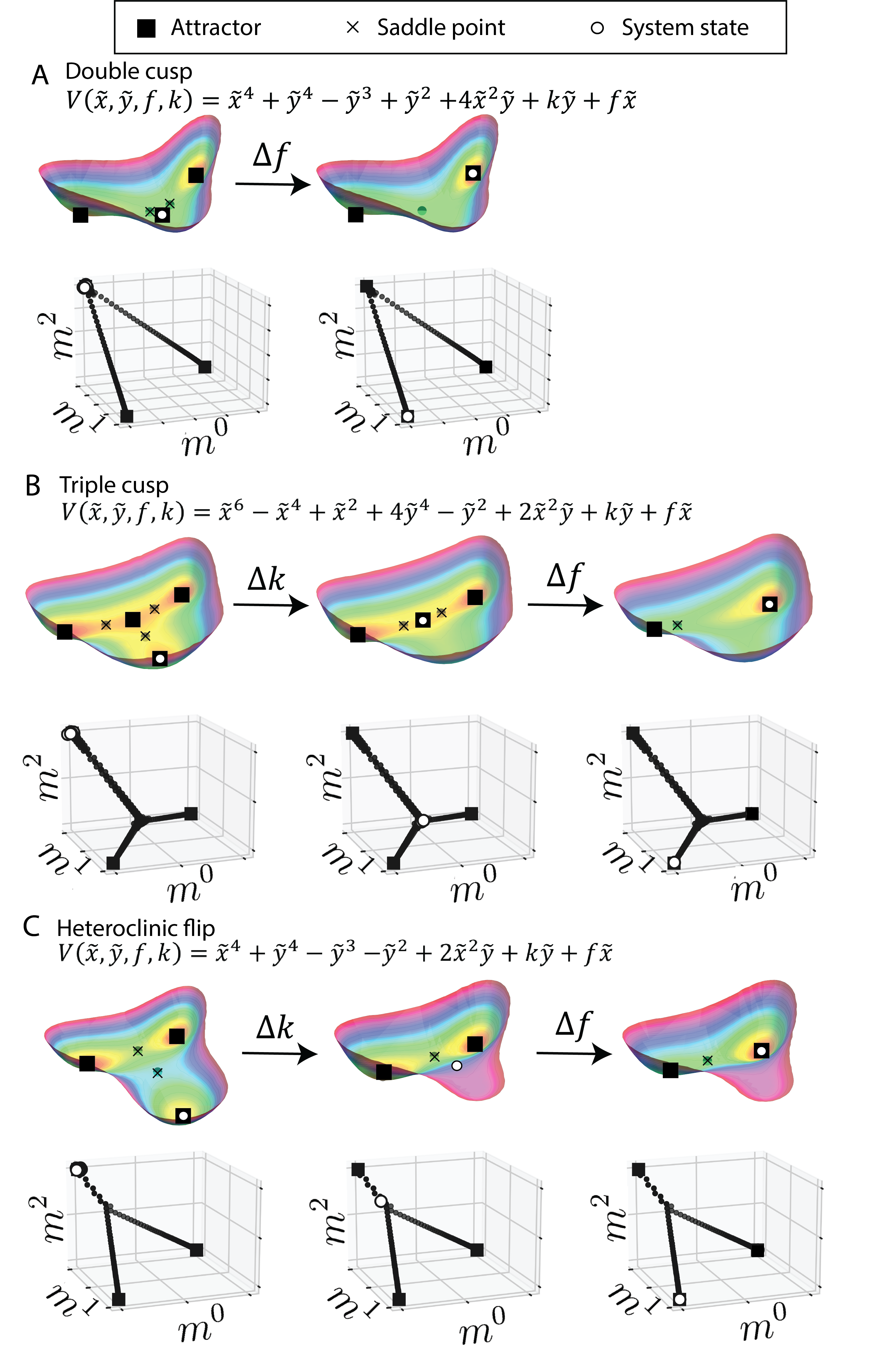}
  \caption{Simulations of cell fate dynamics for three possible landscapes as they undergo bifurcations. Each landscape is colored by depth to emphasize attractor basins. Below each landscape, the full trajectory in cell fate space ($m^0, m^1, m^2$ for cell fates 1, 2, 3) is shown in black, with white circles indicating the cell states corresponding to the white circles in the landscapes above. The equations for each landscape are shown in terms of variables $\tilde{x}, \tilde{y}$ and signaling parameters $f, k$ (see SI section \ref{coordinate transform} for explanation of these coordinates and how they relate to $m^0, m^1, m^2$). For the bifurcation diagrams for each case, see figure \ref{M-bifurcation diagrams}. }
  \label{landscape simulations}
\end{figure}

\begin{figure}
    \centering
    \includegraphics[width=0.9\linewidth]{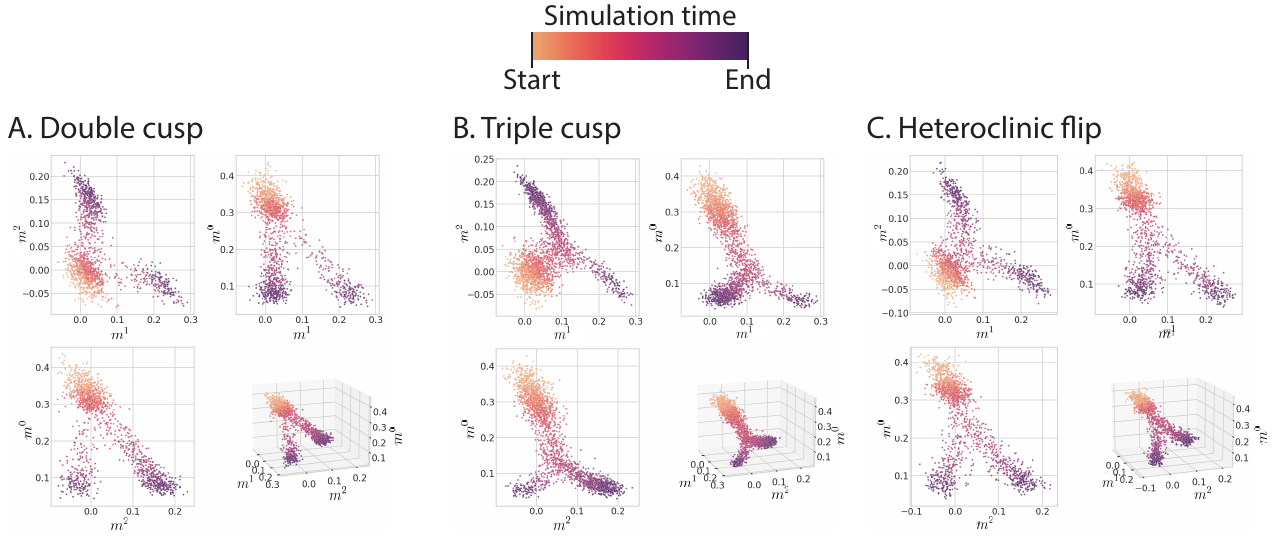}
    \caption{Simulations of cell fate dynamics corresponding to each of the three classes. The simulations are run similarly to Figure \ref{M-signatures}, except noise has been added to imitate measurement noise in scRNA-seq data.}
    \label{fig:noisy classes}
\end{figure}

\subsubsection{\label{coordinate transform} Landscape coordinate transformation} 

\begin{figure}
  \centering
  \includegraphics[width=0.7\textwidth]{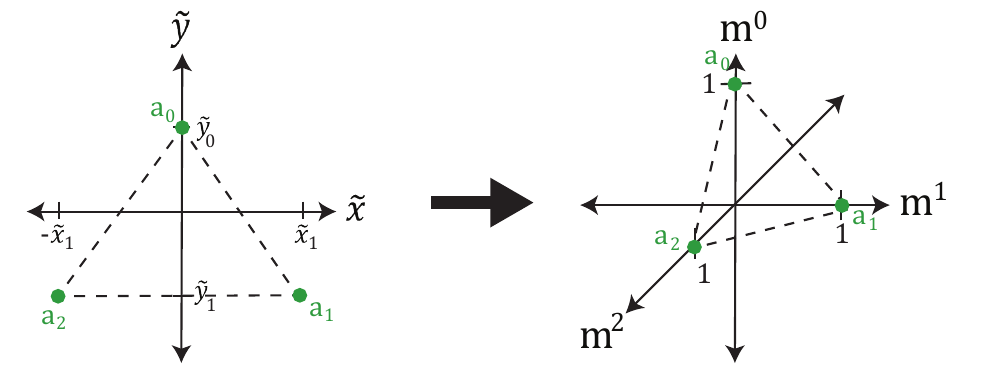}
  \caption{Transforming from normal form coordinates $\tilde{x}, \tilde{y}$ to cell type coordinates $m^0, m^1, m^2$ to align landscape attractors with Hopfield attractors}
\label{coord transform fig}
\end{figure}

The normal forms corresponding to each decision-making classes take the form of a 2-dimensional potential, $V(\tilde{x}, \tilde{y})$. In our model, alignment with each cell type is given its own axis, set by the Hopfield-inspired order parameters $m^{\mu}$. To align the attractors of the landscape $V(\tilde{x}, \tilde{y})$ with the attractors of the Hopfield model, we must transform coordinates from landscape space $\tilde{x}, \tilde{y}$ to cell type space $m^{\mu}$ (see Figure \ref{coord transform fig}). 

In the case of three attractors (points labelled $a_0, a_1,$ and $a_2$), we want to transform from the two dimensions $\tilde{x}, \tilde{y}$ to three dimensions $m^0, m^1, m^2$. For the normal forms of the three bifurcations considered in this paper, the coordinates of these attractors in landscape space $(\tilde{x}, \tilde{y})$ adhere to the following format:

\begin{align*}
    \tilde{a}_0 &= (0, \tilde{y}_0) \\
    \tilde{a}_1 &= (\tilde{x}_1, \tilde{y}_1) \\
    \tilde{a}_2 &= (-\tilde{x}_1, \tilde{y}_1)
\end{align*}

In other words, one attractor is on the $\tilde{y}$-axis and the other two are symmetrically opposed across the $\tilde{y}$-axis. In cell type space $(m^2, m^1, m^0)$, we want the attractors to be on the edges of the 3-simplex:

\begin{align*}
    a^{m}_0 &: (0, 0, 1) \\
    a^{m}_1 &: (0, 1, 0) \\
    a^{m}_2 &: (1, 0, 0)
\end{align*}

To facilitate writing this transformation in matrix format, we will move to four-dimensions such that $(\tilde{x}, \tilde{y})$ coordinates become $(\tilde{x}, \tilde{y}, 0, 1)$ and $(m^2, m^1, m^0)$ coordinates become $(m^2, m^1, m^0, 1)$. The index for summation over this four-dimensional space will be given by $\lambda$. Note that once the transformation is applied, it is simple to move to three-dimensional space ($\lambda = 0, 1, 2, 3 \rightarrow \mu = 0, 1, 2$) by simply truncating the fourth dimension (for equations involving this transform, if $\mu$ is used for summation it implies the 3-dimensional vector is extended to 4 dimensions, with a value of 1 in the fourth dimension and a subsequent summation over $\lambda = 0, 1, 2, 3$).

The 4-by-4 matrix transforming from $(\tilde{x}, \tilde{y}, 0, 1)$ to $(m^2, m^1, m^0, 1)$ is given by $K$. To compose $K$ such that $\vec{a^{m}} = K \vec{\tilde{a}}$, the following transformation steps are necessary:

\begin{enumerate}
    \item Scaling with matrix $S$ such that the triangle formed by $a_0, a_1, a_2$ becomes equilateral with side lengths $\sqrt{2}$. $S = \begin{pmatrix}
        \frac{\sqrt{2}}{2x_1} & 0 & 0 & 0 \\
        0 & \frac{\sqrt{3/2}}{y_0 - y_1} & 0 & 0 \\
        0 & 0 & 1 & 0 \\
        0 & 0 & 0 & 1
    \end{pmatrix}$

    \item Translating the $a_0, a_1, a_2$ triangle with matrix $T_1$ such that the midpoint of the bottom side is a distance $\frac{\sqrt{2}}{2}$ away from the origin. $T1 = 
    \begin{pmatrix}
        1 & 0 & 0 & 0 \\
        0 & 1 & 0 & -\frac{\sqrt{2}}{2} - \frac{y_1 \sqrt{3/2}}{y_0 - y_1} \\
        0 & 0 & 1 & 0 \\
        0 & 0 & 0 & 1
    \end{pmatrix}$

    \item Rotating about the third axis by an angle of $\frac{3 \pi}{4}$ with matrix \[R_1 = 
    \begin{pmatrix}
        \cos\left(\frac{3\pi}{4}\right) & -\sin\left(\frac{3\pi}{4}\right) & 0 & 0 \\
        \sin\left(\frac{3\pi}{4}\right) & \cos\left(\frac{3\pi}{4}\right) & 0 & 0 \\
        0 & 0 & 1 & 0 \\
        0 & 0 & 0 & 1
    \end{pmatrix}\]

    \item Rotating by an angle of $\arcsin\left(\sqrt{\frac{2}{3}}\right)$ about the line that crosses through (1, 0, 0, 1) and (0, 1, 0, 1) with matrix $M = T^{-1} X^{-1} Y^{-1} Z Y X T$. $T, X, Y, Z$ are defined as follows:

    \begin{align*}
        T &= \begin{pmatrix}
            1 & 0 & 0 & -1 \\
            0 & 1 & 0 & 0 \\
            0 & 0 & 1 & 0 \\
            0 & 0 & 0 & 1
            \end{pmatrix} \\
        X &= \begin{pmatrix}
            1 & 0 & 0 & 0 \\
            0 & 0 & -1 & 0 \\
            0 & 1 & 0 & 0 \\
            0 & 0 & 0 & 1
            \end{pmatrix} \\
        Y &= \begin{pmatrix}
            \frac{\sqrt{2}}{2} & 0 & \frac{\sqrt{2}}{2} & 0 \\
            0 & 1 & 0 & 0 \\
            -\frac{\sqrt{2}}{2} & 0 & \frac{\sqrt{2}}{2} & 0 \\
            0 & 0 & 0 & 1
            \end{pmatrix} \\
        Z &= \begin{pmatrix}
            \cos\left(\arcsin\left(\sqrt{\frac{2}{3}}\right)\right) & -\sin\left(\arcsin\left(\sqrt{\frac{2}{3}}\right)\right) & 0 & 0 \\
            \sin\left(\arcsin\left(\sqrt{\frac{2}{3}}\right)\right) & \cos\left(\arcsin\left(\sqrt{\frac{2}{3}}\right)\right) & 0 & 0 \\
            0 & 0 & 1 & 0 \\
            0 & 0 & 0 & 1
            \end{pmatrix}
    \end{align*}

    Multiplying the matrices from all these steps together, we have $K = M R_1 T_1 S$. For transforming from cell-type space $m^{\mu}$ to landscape space $\tilde{x}, \tilde{y}$, we define $H = K^{-1}$. $h_{\mu}^{x, y}$ denote the entries of $H$ such that $\tilde{x} = \sum_{\mu} h_{\mu}^x m^{\mu}$ and $\tilde{y} = \sum_{\mu} h_{\mu}^y m^{\mu}$.
    
\end{enumerate}

\subsubsection{Correlated cell types and invariant transformations}\label{invariant transform}

Cell types are often highly correlated, especially if they have common progenitors. \citet{kanter1987associative} proposed an attractor network storage method which allowed for many highly-correlated attractors to coexist, which was further applied to the modern Hopfield network by \citet{chaudhry2024long}. These generalized order parameters $m^{\mu}$ replace the usual magnetizations $m_{\mu}$. The new order parameters are defined as follows:

\begin{align*}
    A_{\mu \nu} &= \sum_{i} \xi_{\mu i} \xi_{\nu i} \\
    m^{\mu} &= \frac{1}{N} \sum_{\nu j} (A^{-1})^{\mu \nu} \xi_{\nu j} x_j = \sum_{\nu} (A^{-1})^{\mu \nu} m_{\nu}
\end{align*}

These new order parameters are the result of a projection onto the non-orthogonal subspace of cell fates. It is a change of basis with $g_{\mu \nu} = A_{\mu \nu}$ acting as the metric tensor. We adopted Einstein notation to clarify when the metric is applied, with $g^{\mu \nu} = g_{\mu\nu}^{-1}$. The generalized order parameter is a contravariant vector defined with an upper index, $m^{\mu} = \sum_{\mu} g^{\mu \nu} m_{\nu}$, while the classic Hopfield order parameter is a covariant vector with a lower index $m_{\mu}$. The potential $V$ is a scalar, and any dot products used to calculate this scalar must take the metric tensor into account. For example, $V = \vec{m}\cdot \vec{m}$ would become $V = \sum_{\mu} m^{\mu} m_{\mu} = \sum_{\mu \nu} m_{\mu} g^{\mu \nu} m_{\nu}$. 

The conversion from $(\tilde{x}, \tilde{y})$ coordinates to $(m^0, m^1, m^2)$ coordinates is defined in Section~\ref{coordinate transform} as $\tilde{x} = \sum_{\mu} h_{\mu}^x m^{\mu}$ and $\tilde{y} = \sum_{\mu} h_{\mu}^y m^{\mu}$. Using the metric tensor, these are $\tilde{x} = \sum_{\mu \nu} h_{ \mu}^x g^{\mu \nu} m_{\nu}$ and $\tilde{y} = \sum_{\mu \nu} h_{\mu}^y g^{\mu \nu} m_{\nu}$. Thus, the derivative $\frac{\partial V}{\partial m_{\mu}}$ involves the metric tensor as follows:

\begin{align*}
    \frac{\partial V}{\partial m_{\mu}} &= \frac{\partial V}{\partial \tilde{x}} \frac{\partial \tilde{x}}{\partial m_{\mu}} + \frac{\partial V}{\partial \tilde{y}} \frac{\partial \tilde{y}}{\partial m_{\mu}} \\
    &= \frac{\partial V}{\partial \tilde{x}} (\sum_{\lambda} h_{ \lambda}^x g^{\lambda \mu}) + \frac{\partial V}{\partial \tilde{y}} (\sum_{\lambda} h_{\lambda}^y g^{\lambda \mu})
\end{align*}

For clarity, since $h^{x,y}$ are actually 4-dimensional vectors, the notation used above corresponds to the following: in the products $\sum_{\mu \nu} h_{\mu}^{x,y} g^{\mu \nu} m_{\mu}$ and $\sum_{\lambda} h_{\lambda}^{x, y} g^{\lambda \mu}$, $g^{\mu \nu}$ and $m^{\mu}$ are moved to 4-dimensional space with the addition of a homogenous coordinate. These calculations are written out explicitly below.

\begin{align*}
    \sum_{\mu \nu} h_{\mu}^{x, y} g^{\mu \nu} m_{\nu} \rightarrow h^{x, y} \cdot g \cdot m &= \begin{pmatrix}
        h_{0}^{x,y} \\
        h_{1}^{x,y} \\
        h_{2}^{x,y} \\
        h_{3}^{x,y}
    \end{pmatrix} \begin{pmatrix}
        g^{00} & g^{01} & g^{02} & 0 \\
        g^{10} & g^{11} & g^{12} & 0 \\
        g^{20} & g^{21} & g^{22} & 0 \\
        0 & 0 & 0 & 1
    \end{pmatrix} \begin{pmatrix}
        m^0 \\
        m^1 \\
        m^2 \\
        1
    \end{pmatrix}
\end{align*}

The derivative $\frac{\partial V}{ m_{\mu}}$ contains the terms $\sum_{\lambda} h_{\lambda}^{x, y} g^{\lambda \mu}$. In this case, $g^{\lambda \mu}$ is a 4-by-4 matrix as written above for the product, and after the product is calculated the fourth dimension is truncated so that the index $\mu$ refers once again to cell fates $0, 1, 2$. We can write this explicitly using $\alpha$ and $\beta$ to refer to the 4-dimensional space (three cell fate dimensions plus the extra dimension for transforming coordinates):

\begin{align*}
    \sum_{\lambda} h_{\lambda}^{x, y} g^{\lambda \mu} \rightarrow \sum_{\alpha} h_{\alpha}^{x, y} g^{\alpha \beta} &= (h^{x,y} \cdot g)_{\beta}
\end{align*}

To bring this back to three coordinates for $\frac{\partial V}{\partial m_{\mu}}$, we simply take $\beta = 0, 1, 2$ and ignore $\beta = 3$.

\subsubsection{Potentials over time}\label{energy landscapes}
It's well-known that the Hopfield model has an associated energy landscape which acts as the Lyapunov function of the system \cite{krotov2020large}. In other words, the update rule of the Hopfield model causes this energy function to decrease. For the modern Hopfield model, the energy takes the following form:
\begin{align*}
    E = \frac{1}{2N} \sum_{i=1}^{N} x_i^2 - \log(\sum_{\mu} \exp(\beta m^{\mu}))
\end{align*}
It's unclear what the corresponding Lyapunov function would be for the modified Hopfield model we introduce in this paper. However, we can compare how the intrinsic energy (defined above) and the extrinsic energy (the potential $V$ we insert into the system) change over time. We refer to the first term in the intrinsic energy as the "kinetic" term and the second term as the "intrinsic" term.

\begin{figure}
    \centering
    \includegraphics[width=0.75\linewidth]{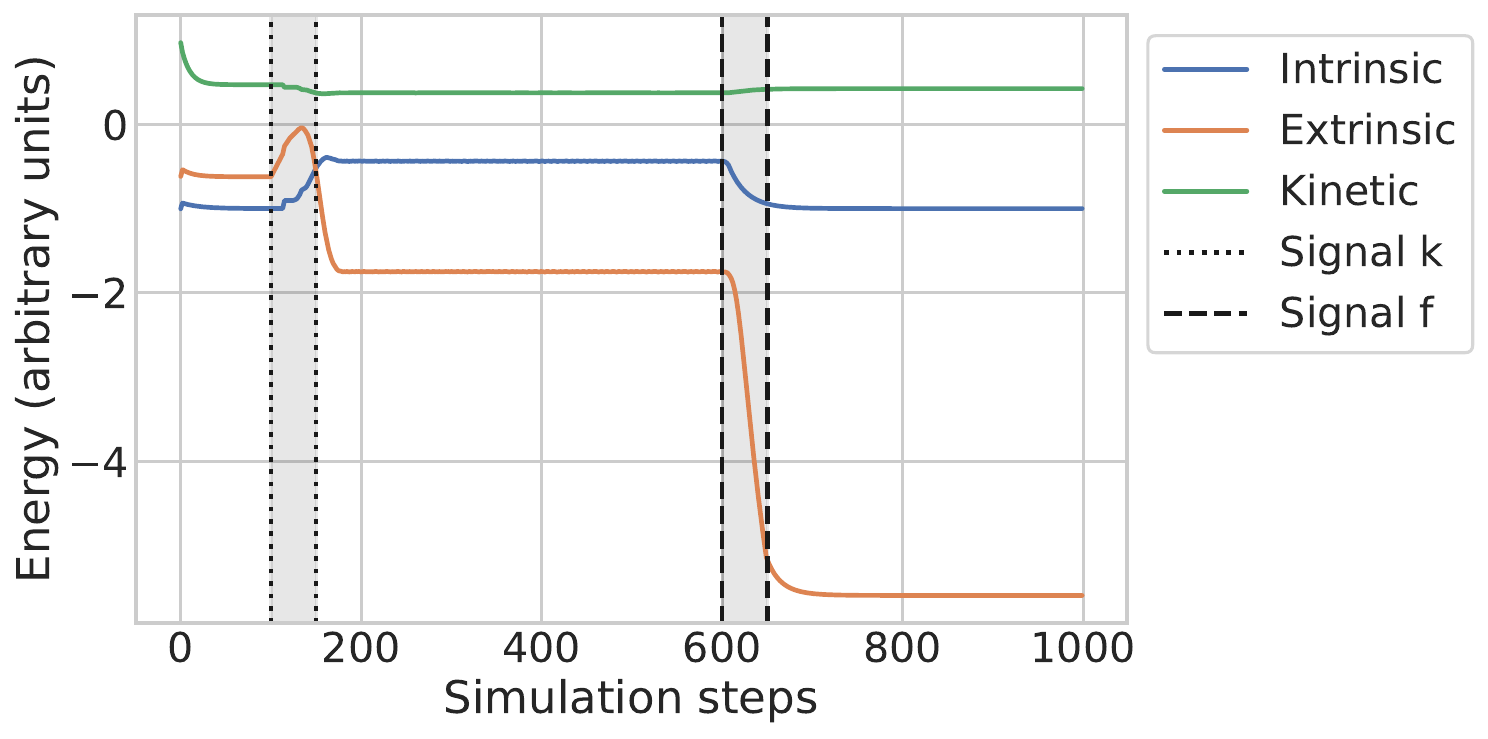}
    \caption{The Hopfield energy and the imposed potential change as the simulation progresses. The imposed potential decreases while the Hopfield energy has a temporary increase. "Intrinsic" refers to the second term of $E$, the Hopfield energy. "Extrinsic" refers to the imposed potential $V$. "Kinetic" refers to the first term of the Hopfield energy $E$ . The units are arbitrary because, since the actual form of the Lyapunov function of the modified Hopfield model is not known, the scale has no meaning.}
    \label{fig:energy update}
\end{figure}

In Figure \ref{fig:energy update}, we show how these different energies change over simulation time for the triple cusp landscape. This is the same simulation as shown at the top of Figure \ref{update schedule}: the first change in signalling (signal $f$) causes the cell to leave the first attractor basin, and the second change in signalling (signal $k$) causes the cell to leave the intermediate state and enter the final basin corresponding to a mature cell type. As shown in the figure, although the intrinsic energy increases temporarily when the cell enters a state not corresponding to one of the Hopfield stored patterns, the extrinsic energy decreases throughout the simulation. 

\subsubsection{Spatial patterning}

In Section \ref{M-spatial}, to simulate cells with a spatial pattern of alternating cell fates, we model the system as a line of cells. Each cell experiences a landscape that is biased towards one fate or another by its neighboring cells. To identify the effects of different landscapes, we simulated both the double cusp and triple cusp landscapes. Instead of manually changing the signals $f$ and $k$ biasing the landscapes towards one fate or another, $f$ is a function of the neighboring cell types. The signaling parameter $f_i$ affects the $i$th cell in a one-dimensional chain of 20 cells. $f_i$ depends on the cell fate identity of the cell's left neighbor $\vec{m}({i-1})$ and the identity of the right neighbor $\vec{m}({i+1})$. To model lateral inhibition, where ciliated cells will push their neighbors towards club fates and vice versa, we chose $f_i$ to be a sigmoid dependent on the difference between ciliated a club identities:
\begin{align*}
    f_i &= -A (\frac{2}{1 + e^{a s_i}} - 1) \\
    s_i &= \max_{(i+1, i-1)} |m^{ciliated} - m^{club}|
\end{align*}

\begin{figure}
    \centering
    \includegraphics[width=0.75\linewidth]{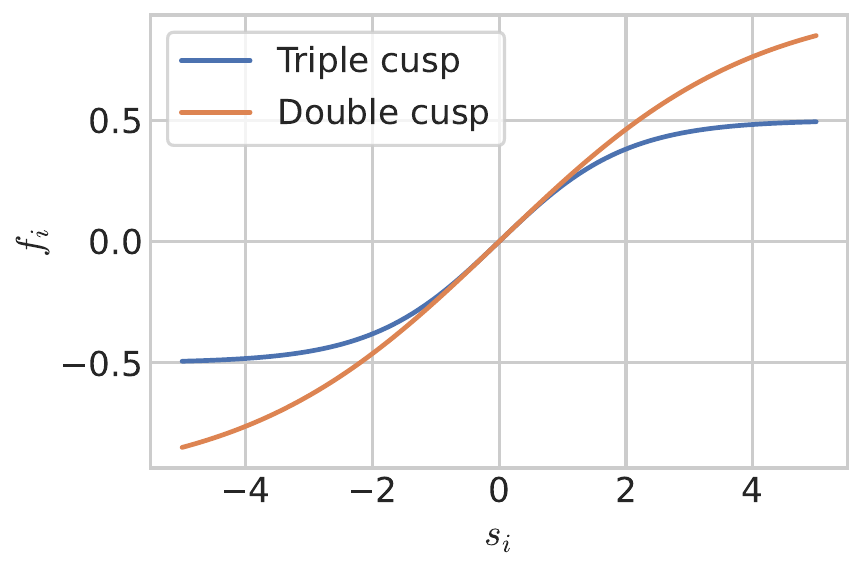}
    \caption{The signalling parameter $f_i$ for cell $i$ in a chain of cells, as a function of signalling input $s_i$. In simulations of the triple and double cusp, the sensitivities $a$ were chosen to be the same although the amplitudes $A$ are different due to differences in the signalling parameter necessary for bifurcations in either landscape.}
    \label{fig:signal param}
\end{figure}

If a neighboring cell is more ciliated than club, $f_i > 0$, the signal biases the landscape towards the club fate. Likewise, if $m^{club} > m^{ciliated}$, $f_i < 0$ and the landscape of the $i$th cell is biased towards the ciliated fate. The signaling input $s_i$ takes the maximum over the two neighboring cells, so that the neighbor with a stronger identity determines the $i$th cell's bias. To avoid differences in signal sensitivity between neighbors, we chose $a$ to be the same value across landscapes while $A$ varied due to differences in basin depth (as shown in Figure \ref{fig:signal param}).

\subsection{Simulation details}

The full details of the simulation can be found in the accompanying Python Jupyter notebooks. This section describes the structure of the code as well as some important aspects of the simulations performed to produce the figures in the main text.

Simulating a cell fate transition begins by instantiating an object of the\ \codeword{AttractorNetwork} class. The class simulates dynamics for the case of three primary attractor states. It works by storing the state of the network, $x_i$, and simulating the dynamics $\frac{dx_i}{dt}$ for specified landscape, signals, and attractor states. This class is structured to be modular, so it can take any suitable landscape as input. It requires the gradient of the landscape $(\frac{dV}{d\tilde{x}}, \frac{dV}{d\tilde{y}})$ and the attractor coordinates $\tilde{a}_0,\tilde{a}_1,\tilde{a}_2$ in $(\tilde{x},\tilde{y})$ space. The gradient of the landscape is a function of the signaling parameters $(k,f)$, which control the bifurcations.  The initial and final values of the signaling parameters $(k_{i,f}, f_{i,f})$ and the signal timing of the signals are specified at the start of the simulation. It uses this information, along with the attractor states $\xi_{\mu i}$, to calculate $h_{ \lambda}^{x,y}$ and $g^{\lambda \mu}$, which are then used to calculate the update step:
\begin{align*}
x_i(t+1) &= x_i(t) + \frac{dx_i}{dt}\\
\tau \frac{dx_i}{dt} &=  \sum_{\mu} \xi_{\mu i}  \sigma^\mu(m^{\mu}(t) -\frac{\partial V}{\partial \tilde{x}} (\sum_{\lambda} h_{ \lambda}^x g^{\lambda \mu}) - \frac{\partial V}{\partial \tilde{y}} (\sum_{\lambda} h_{\lambda}^y g^{\lambda \mu}))-x_i(t)\\
\end{align*}
$\tau$ sets the speed of the dynamics. The result of varying $\tau$ or setting an asynchronous update schedule are discussed in Section \ref{update schedule}. The signal timing is discussed in Section \ref{signal dynamics}.

The simulations were run in gene expression space ($x_i$) but the plots are shown are in cell fate space ($m_{\mu}$). The cell fate coordinates were calculated using the attractor states $\xi_{\mu i}$. For figures \ref{M-signatures}, \ref{fig:noisy classes}, the stored patterns used in the simulation were three scRNA-seq samples of alveolar type 1, alveolar type 2, and early epithelium cells. The former two are from \citet{herriges2023durable} while the latter is from \citet{negretti2021single}. 17,970 genes were simulated. 

\subsubsection{Signaling dynamics}\label{signal dynamics}
For Figure \ref{M-bifurcation diagrams}, the potentials were plotted in Mathematica using the equations shown. The values for the signaling parameters $(k,f)$ were varied as follows. For the landscape with a triple cusp, the system starts with $k=0, f=0$. The signaling parameter $k$ is increased from 0 to 0.3, destabilizing the attractor in which the system started. Then, $f$ is changed from 0 to 0.3, causing the intermediate attractor to disappear and tilting the landscape towards one of the final fates. In the landscape with a double cusp bifurcation, the parameters start at $k=0.15, f=0$.  Parameter $k$ is kept at 0.15, and $f$ is changed from 0 to 0.1. This destabilizes the initial attractor and pushes the cell towards one of the final cell fates. The landscape with a heteroclinic flip bifurcation starts with $k=0.5, f=0$. Signal $k$ is shifted from 0.5 to 2 to destabilize the initial attractor, and before that bifurcation is fully complete, $f$ is changed from 0 to 0.5 to tilt the landscape towards the cell fate on the right-hand side. 

In the simulated trajectories shown in Figures \ref{M-signatures} and \ref{fig:noisy classes}, the signaling parameters $f, k$ were varied manually in short time windows to reflect changes in the landscape caused by received signals. 10 trials were run for each landscape class. Not many trials were needed since the dynamics are so consistent (besides the added noise, see Section \ref{noise}) and there were enough time steps to produce a lot of data. Generally, the timing of signals was chosen to emphasize the features unique to each class. If the signals are changed very quickly, the system quickly converges to its final point. If the signals are changed slowly enough, one can find cells along the entire trajectory. 

Start and end times, as well as start and end values, are specified for each parameter when the \codeword{AttractorNetwork} class is instantiated. $t^f_{s}, t^f_{e}$ are the start and end times for parameter $f$ while $t^k_s, t^k_e$ are the start and end times for parameter $k$. $(f_{s}, f_e), (k_{s}, k_{e})$ indicate the start and end values for each of the parameters. Within the time window specified by the start and end times for each signal, the parameter was changed as follows:
\begin{align*}
    f(t+1) &= f(t) + \frac{f_e - f_s}{t_e^f - t_s^f} \\
    k(t+1) &= k(t) + \frac{k_e - k_s}{t_e^k - t_s^k}
\end{align*}

For the triple cusp, the two signals represent two stages of differentiation. The first stage is the transition from the initial state to the intermediate state, and the second stage is the specification into one of the final two fates. Each simulation was run for 1000 time steps. The signaling schedule was chosen so that the cell transitioned to the intermediate attractor first, before receiving signals that push it to its final fate. The initial signaling parameters were $k_s = 0, f_s =0$. From time steps 0 to 400, $k$ was gradually changed from $k_s = 0$ to $k_e = 0.3$. This change represents an initial signal that pushes the cell out of its initial attractor state. Then, from steps 600 to 700, $f$ was changed from $f_s = 0$ to $f_e = \pm 0.3$. This represents a signal that pushes the cell to one of the final two states. The sign of the signal is chosen randomly for each trial in order to show routes to both final fates. The first signal was varied over a longer stretch of time than the second to show cells along the first leg of the path. Also, there is a time gap between the signals of 200 time steps. The length of this gap determines how long cells stay in the intermediate attractor and therefore how prominent that cluster is in the plots. If the time between signals is too short, the intermediate attractor is difficult to detect. 

In the double cusp, changing either signal can destabilize the attractor. $f$ represents the bias towards one of the two final fates. If $k$ is changed while $f$ remains zero, the initial attractor basin becomes a saddle point, and the cell has equal probability to end up in either final attractor if it is perfectly positioned on the saddle point. The bifurcation from changing $f$ is more interesting than the one from changing $k$ because the latter only involves a random choice rather than a decision, so we only varied $f$ for this bifurcation. The simulation ran for 750 time steps per trial. At the beginning of the simulation, $f=0$ and $k=0.15$. For the duration of the entire simulation, $f$ is gradually changed to $\pm 0.1$. As for the triple cusp, the sign of this signaling parameter was randomly chosen for each trial in order to cover both possible paths in the transition. 

In the heteroclinic flip landscape, the first signal pushes the cell out of the initial attractor towards a saddle point and the second signal biases the cell towards one of the final two fates. In the simulations, this system starts with $f_s = 0, k_s = 0.5$. Changing $k$ destabilizes the initial attractor. Once the initial attractor is destabilized, the system is very sensitive to changes in $f$ because it is poised at a saddle point between the final two fates. For this class, we chose to have an overlap in the time windows of signals. This is to avoid a situation where the cell moves out of the initial attractor towards the saddle point, starts moving towards one of the final fates and then switches directions due to a change in $f$. Each simulation was run for 750 steps. From steps 100 to 500, $k$ was changed from $k_s = 0.5$ to $k_e = 2$. From steps 450 to 600, $f$ was changed from $f_s = 0$ to $f_e = \pm 0.5$. Similarly to the other two landscapes, the sign of $f_e$ was chosen randomly for each trial in order to show trajectories to both final fates.

\subsubsection{Inverse temperature}\label{inverse temp}
For all simulations except when explicitly stated otherwise, the inverse temperature $\beta$ in the softmax of the update rule was set to $\beta={2N}$, where $N$ is the number of genes simulated. For the modern Hopfield network to converge to the correct pattern, it is necessary that the temperature is low enough such that $e^{\beta} \gg P$, where $P$ is the number of stored patterns. This can be seen by considering the softmax in the update rule when $m^{\mu} = 1, m^{\nu \neq \mu} = 0$:

\begin{align*}
    \sigma (\beta m_{\mu}) &= \frac{e^{\beta}}{e^{\beta} + e^0 + e^0 +...}\\
     &= \frac{1}{1 + P e^{-\beta}} \\
     \sigma (\beta m_{\nu \neq \mu}) &= \frac{e^{0}}{e^{\beta} + e^0 + e^0 +...}\\
     &= \frac{1}{e^{\beta} + P}
\end{align*}
In this case, the update rule should keep the state static, so $\sigma(\beta m^{\mu})=1$ and $\sigma(\beta m^{\nu \neq \mu}) =0$. This condition is met for $e^{\beta} \gg P$. 

\subsubsection{Noise}\label{noise}

Noise was added to the simulation in Figures \ref{fig:noisy classes} and \ref{M-signatures} to imitate the noise in scRNA-seq and facilitate comparison between simulated trajectories and data. The simulations were initiated at the initial attractor state with noise added to the gene expression: $x_i (t=0) = \xi_{0 i} + z_i$. The noise $z_i$ for each gene $i$ is drawn from identical, independent normal distributions. Modern Hopfield networks are excellent at converging and de-noising, so after one step the simulation converges to the noise-less trajectory. To maintain noise along the path, we added noise to the recorded variables. Specifically, normally-distributed noise was added to the observed $x_i$ at each step, and a fraction of counts were randomly set to zero to imitate the high levels of dropout in scRNA-seq \cite{hicks2018missing}. Additionally, a large percent of the simulated cells have not been included in the plot, to reflect the fact that scRNA-seq experiments can't capture every cell at every step of the process. The addition of noise makes it harder to distinguish between classes, but contour plots (as in Figure \ref{M-signatures}) help highlight important features.

\subsubsection{Robustness to update schedule}
\begin{figure}
  \centering
  \includegraphics[width=0.6\textwidth]{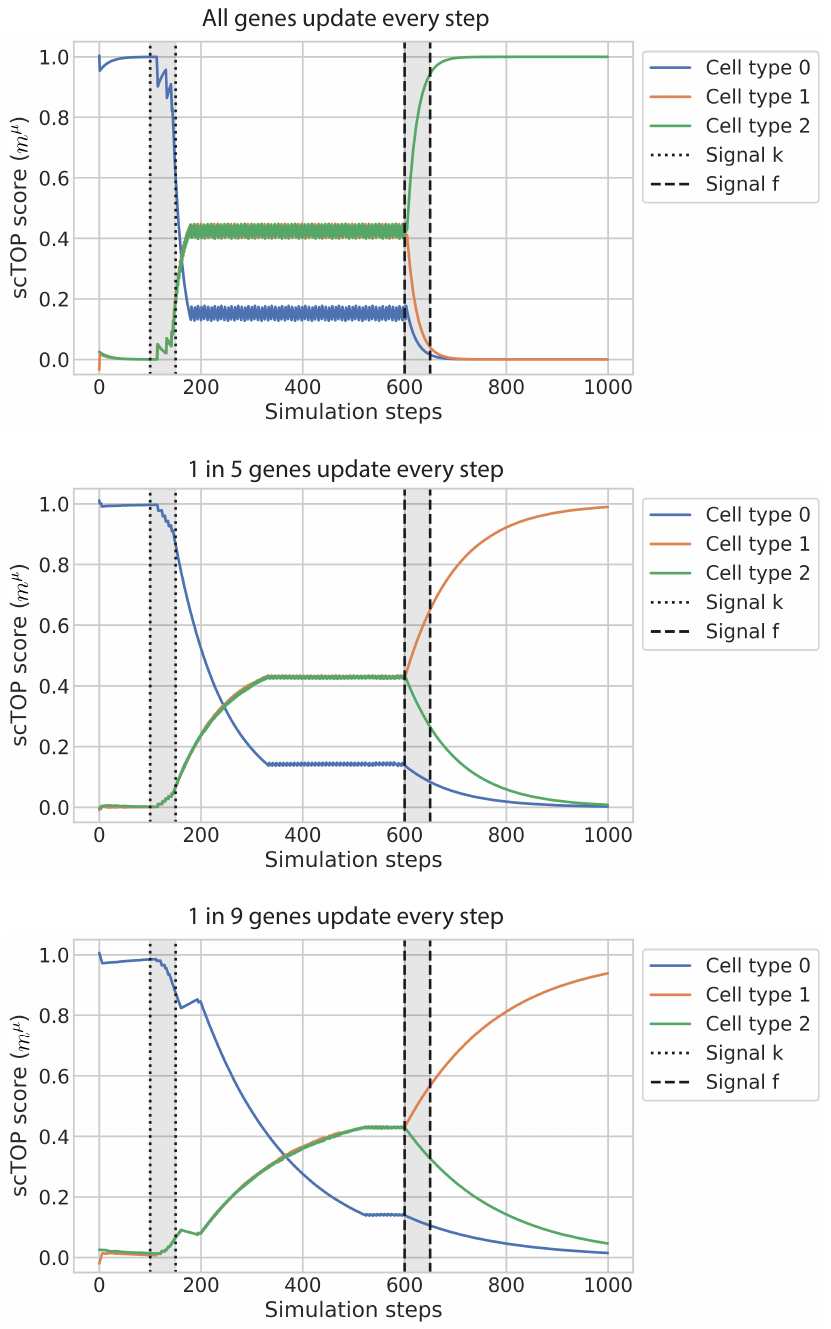}
  \caption{The dynamical system is robust to changes in update schedule. All three plots show the change in cell type alignment in a triple cusp bifurcation over the course of simulation time, where the changes in signal $\Delta f$ and $\Delta k$ occur in the indicated intervals. In each plots, the fraction of genes randomly chosen to be updated at each simulation step is varied, ranging from all genes (top) to 1/5 genes (middle) to 1/9 genes (bottom). The basic behavior remains the same regardless of how many genes are updated, although the time scales vary.}
  \label{update schedule}
\end{figure}

In our model, all genes are updated simultaneously. However, this does not reflect biological reality. To reflect the asynchronous nature of gene regulation, we can randomly update some genes at each simulation step and not others. The model is robust to such changes, and still converges to the expected basins. This is illustrated in Figure \ref{update schedule}, where a different percentage of genes are randomly chosen to be updated in the simulation of a triple cusp bifurcation. The only difference caused by this change in update schedule is a longer time to converge to the attractor basin. This timing only matters relative to the timing of signaling; as long as it occurs on faster timescales than signaling, the basic behavior of the system is not affected.

\subsection{Limitations of the model}\label{limitations}
Because decision-making classes are describe general characteristics, there are many facets of our simulations that were not constrained by choice of bifurcation. In the future, with more analysis using scTOP scores to plot data on the order parameter axes, it may be possible to place more constraints on the simulations. For this paper, we tried to make choices that minimized additional assumptions. For example, in transforming between coordinates, we chose to align the attractor coordinates in $x-y$ space with the simplex in $\vec{m}$ space.

The paths of transition predicted by our model are highly dependent on the choice of potential, control parameters, and other mathematical details. Although the decision-making classes are generic, the choices we made in choosing the forms of the potentials and in transforming between the 2-dimensional $V(x, y)$ and the three-dimensional $V(m^0, m^1, m^2)$ are not generic. With the changes of control parameters, the attractor basins shift slightly in position, which is not necessarily reflective of actual observations. With more scTOP analyses, it will be more and more possible to fit these various parameters and make more informed decisions.

Another limitation is the fact that it is not possible to fully determine the landscape. This is an inherent aspect of coarse-grained models. Not only are there infinite possibilities for connecting decision-making classes, but different landscapes can produce the exact same trajectories in certain regimes (see section \ref{choice of potential}). It is easier for the model to eliminate particular classes of bifurcations than definitively identify the exact class. 

\begin{figure}
    \centering
    \includegraphics[width=0.8\linewidth]{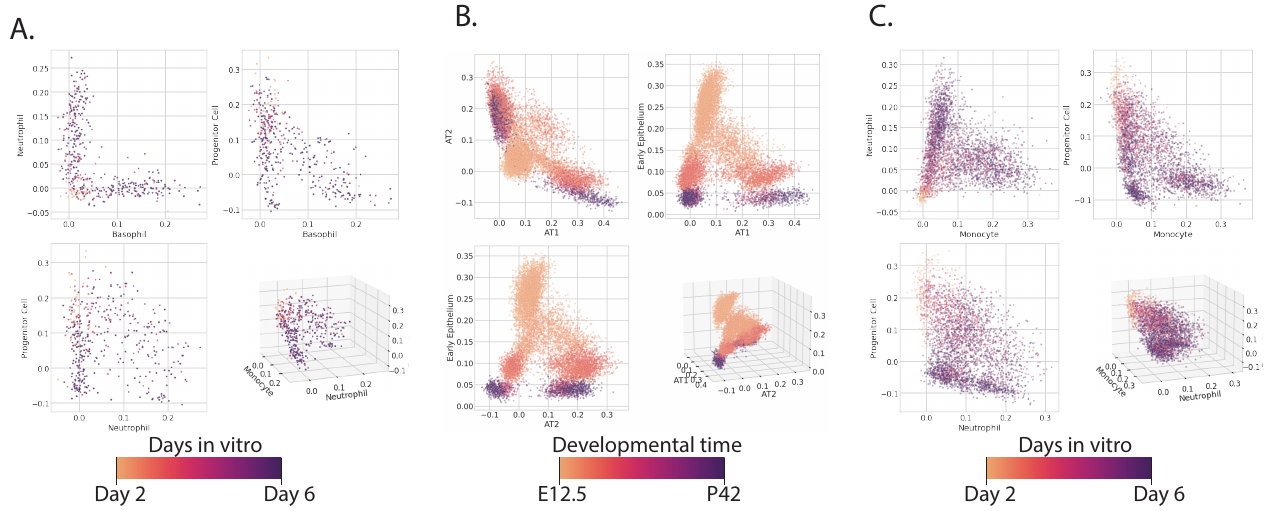}
    \caption{The same data as in Figure \ref{M-signatures in data}, shown as a scatter plot colored by time point. The axes correspond to the two mature fates relevant to each experiment as well as a progenitor cell type, to show the maturation from an early state to a later one.}
    \label{fig:data scatter}
\end{figure}
\bibliography{export-data.bib}